\documentclass[aps,pre,showpacs,groupedaddress]{revtex4}
\usepackage{graphicx}

\newcommand{\be}{\begin{equation}}
  \newcommand{\ee}{\end{equation}}

\newcommand{\rd}{{\rm d}}

\newcommand{\cP}{{\cal P}}

\newcommand{\Bin}{\left( \begin{array}{c} k \\[-0.5ex] m \end{array}  \right)}

\newcommand{\half}{\textstyle { \frac{1}{2}}\,}

\def\simg{\mathrel{\mathchoice {\vcenter{\offinterlineskip\halign{\hfil
          $\displaystyle##$\hfil\cr>\cr\sim\cr}}}
    {\vcenter{\offinterlineskip\halign{\hfil$\textstyle##$\hfil\cr
          >\cr\sim\cr}}}
    {\vcenter{\offinterlineskip\halign{\hfil$\scriptstyle##$\hfil\cr
          >\cr\sim\cr}}}
    {\vcenter{\offinterlineskip\halign{\hfil$\scriptscriptstyle##$\hfil\cr
          >\cr\sim\cr}}}}}

\begin{document}


\title{Activity patterns on random scale-free networks: Global dynamics arising from local 
  majority rules\footnote{Citation information: H. Zhou and R. Lipowsky, J. Stat. Mech. (2007) P01009.}}

\author{Haijun Zhou$^{1}$ and Reinhard Lipowsky$^{2}$}

\affiliation{$^1$Institute of Theoretical Physics, the Chinese Academy of Sciences, Beijing 100080, 
  China}

\affiliation{$^2$Max-Planck-Institute of Colloids and Interfaces, Potsdam 14424, Germany}

\date{\today}

\begin{abstract}
  Activity or spin patterns on random scale-free network are studied 
  by mean field analysis and computer simulations.  These activity patterns
  evolve in time according to local majority-rule dynamics which is implemented 
  using (i) parallel or synchronous updating and (ii) random sequential or asynchronous
  updating.   Our mean-field   calculations predict that  the relaxation processes of 
  disordered activity patterns become much more efficient as the
  scaling exponent $\gamma$ of the scale-free degree distribution changes from $\gamma >5/2$ to $\gamma < 5/2$. 
  For $\gamma > 5/2$, the corresponding decay times increase as 
   $\ln(N)$ with increasing network size $N$ whereas they are  independent of $N$
   for $\gamma < 5/2$.  In order to check these mean field 
   predictions, extensive   simulations of the pattern dynamics have been performed 
   using two different ensembles of random scale-free networks: (A) multi-networks as generated by 
   the configuration method, which typically leads to many self-connections and multiple edges, 
   and (B) simple-networks without self-connections and multiple edges. We find that the mean field 
   predictions are confirmed (i) for random sequential updating of multi-networks and (ii) for both 
   parallel and random sequential updating of simple-networks with 
   $\gamma = 2.25$ and $\gamma = 2.6$. For $\gamma = 2.4$,   the   data for the simple-networks 
  seem to be consistent with mean field theory as well whereas we cannot draw a definite conclusion 
  from the simulation data for the multi-networks. The latter difficulty  can be
  understood in terms of an effective scaling exponent $\gamma_{\rm eff} = \gamma_{\rm eff}  (\gamma, 
  N)$  for   multi-networks. This effective exponent is determined by removing  all self-connections 
  and multiple edges;  it  satisfies  $\gamma_{\rm eff} \geq \gamma$ and  decreases towards $\gamma$
  with increasing network size $N$. For $\gamma = 2.4$, we find  $\gamma_{\rm eff}  \simg 5/2$
  up to $N = 2^{17}$. 
\end{abstract}


\maketitle

\section{Introduction}
\label{sec:introduc}

Both the overall topology and the local structure of  complex networks influence the efficiency
as well as the  robustness or sensitivity  of  dynamic processes occurring on these
networks. Furthermore, the local network architecture often evolves with time in order
to optimize certain network properties. 
This interplay of structure and dynamics eventually leads to the formation of
highly optimized network topologies for specific dynamic processes.

Understanding the underlying mechanism for the optimization of networks via 
evolution is an important but challenging problem of current network research.  
There are two major difficulties. First, the optimization of different
dynamic processes may require different  network architectures. For example, we
would not be surprised to find out that  the architectures of  car traffic  networks  and 
electronic networks on   micro-processor chips are rather different.  
Therefore in order to study network evolution, we should focus on a particular dynamic process.
Second, the evolution process is usually governed by some more or less complex 
growth rules
which are coupled in some unknown way to the 
 dynamic process under consideration. 
  
 In this article, we use a somewhat different approach and  study the influence of 
network architectures on relatively simple
dynamic processes. By performing such an analysis, we are able to identify 
those aspects of the network structure that are important for the dynamic processes.  Such a
theoretical investigation should also be useful to analyze the available empirical data 
about complex networks in terms of optimization to design improved network structrues for a 
particular dynamic process.

A network can be represented by a graph of vertices (nodes) and edges (links). Each edge of the 
network connects two vertices of the network. The vertex degree  $k$  is equal to the
total number of edges that are connected to this vertex.
Recent structural studies on complex networks
\cite{Albert-Barabasi-2002,Dorogovtsev-Mendes-2002,Newman-2003}
show that  the topology of  real-world networks often has the so-called
scale-free property \cite{Barabasi-Albert-1999}. This means that the total number $N( k )$  of
vertices  with degree $k$ scales with $k$ as a
power-law, $N( k ) \sim k^{-\gamma}$, with   the  decay  or 
scaling exponent $\gamma$. Such a power-law scaling of the degree
distribution appears if the network grows via a preferential attachment mechanism
\cite{Barabasi-Albert-1999}.  It is also plausible that the scale-free scaling is a
result of  network evolution and optimization, but so far there are only a 
few theoretical  studies that have addressed this issue (e.g., 
Refs.~\cite{Valverde-etal-2002,FerrerICancho-Sole-2003,Variano-etal-2004,Wang-etal-2005}).

In the following, we will consider
 the influence of scale-free random networks on the activity patterns of binary or Ising-like
 variables which are governed by    local  majority-rule dynamics. 
We focus on scale-free networks because they are abundant 
in our biosphere. We study  local majority-rule dynamics 
 because it is simple and  frequently encountered in the real world, 
and also because it can be readily extended to a large  class of more complicated 
dynamic processes.

We perform  both analytical mean-field calculations and extensive
numerical simulations on these model systems. Our mean-field  calculations 
predict that, when the    scaling exponent $\gamma$ of the substrate network
changes from $\gamma >5/2$ to $\gamma < 5/2$, a qualitative improvement 
in the efficiency of the dynamic process can be achieved, namely the typical relaxation
time changes from being proportional to $\ln(N)$ to being independent of $N$, 
where $N$ is the total number of nodes in the network.
In the simulations,  we consider two different methods to generate ensembles of 
random scale-free networks: (i) multi-networks, which are generated using the 
so-called configuration model and 
 contain some multiple edges and self-connections; and (ii) simple-networks, which involve
 some random reshuffling of edges and do not contain any multiple edges or self-connections. 
Our mean-field results are quantitatively confirmed by the simulations for simple-networks
whereas the behavior on multi-networks exhibits strong finite-size corrections arising from the 
presence of multiple edges and self-connections. 
We use both parallel (or synchronous) and random sequential (or asynchronous) 
updating and find that both updating methods give very similar results for 
simple-networks whereas they lead to somewhat different results for multi-networks.  
The present study extends our previous work in 
Ref.~\cite{Zhou-Lipowsky-2005} in which we focussed on parallel updating for multi-networks. 

Our mean-field theory can be extended to related dynamic processes. Instead of Ising-like
variables, we have  studied Potts-like variables which can attain three or more different states. 
We have also  considered more complex dynamic rules for which one applies the local
 majority rule to each spin variable with a certain probaility 
say ${\cal P}$, and another dynamic rule such as random Boolean dynamics with probability 
$1-{\cal P}$. Likewise, we have extended our  analysis to majority rule dynamics on 
directed networks and to Hopfield models on scale-free networks \cite{Zhou-Lipowsky-2005}.
In all cases, the mean-field theorey predicts  that the dynamic behavior changes qualitatively as 
one crosses the boundary 
value $\gamma = 5/2$.  It is of interest to note  that many  scale-free networks
are characterized by scaling exponents $\gamma$ which fall
into the narrow range $2 <\gamma \leq  5/2$ as observed in Ref.~\cite{Aldana-Cluzel-2003}. 
Indeed, table II of Ref. \cite{Albert-Barabasi-2002} contains
a list of ten scale-free networks with $2 < \gamma < 5/2$, one 
with $\gamma = 2.5$, and only three with $2.5 < \gamma < 3$.

This paper is organized as follows. We first describe our model system in
more detail in Sec.~\ref{sec:model}. Then  our mean-field 
analysis is presented in Sec.~\ref{sec:analytic}. Simulation results of local majority-rule dynamics 
are discussed in Sec.~\ref{sec:simulation_MRN} for random multi-networks with  multiple edges and self-connections,  and in
Sec.~\ref{sec:simulation_SRN} for random simple-networks  without any such connections. 
The simulation results for   simple-networks are in complete agreement with the mean field 
predictions whereas we find only  partial agreement between    
 the simulation results for  multi-networks and mean field theory. We then argue in 
 Sec.~\ref{sec:effective_scaling} that
the remaining  discrepancies can be understood if one characterizes the multi-networks by 
an  effective $N$--dependent scaling exponent $\gamma_{\rm eff}\geq \gamma $. 
Finally we conclude our work in Sec.~\ref{sec:conclusion} with a summary and an  outlook on related problems.

\section{Scale-free networks and majority-rule dynamics}
\label{sec:model}

In the following, 
we study local majority-rule dynamics on scale-free random networks. In order to do so, 
we first describe the ensemble of random networks used in our study and then
define the dynamic rules. 

\subsection{Scale-free degree sequences}
\label{eq:model_degree}

We consider an ensemble of  random networks. Each network consists  of $N$ vertices   and edges that connect pairs of vertices.
Each vertex $i$ with $i=1,2,\ldots,N$  is connected to
$k_i \geq k_0$   edges where  $k_0$ represents the minimal value of the vertex degrees $k_i$. 
As in Ref.~\cite{Zhou-Lipowsky-2005}, 
we will focus on $k_0 \geq 2$ for which the network has no `dangling
ends' and consists of many cycles. 
We  consider an ensemble of random networks that provide realizations of the degree distribution 
\be
  \label{eq:sf_degree_profile}
  P( k_i ) \propto  k_i^{-\gamma}  \quad {\rm for} \quad  k_i \geq k_0  
\ee
which is characterized by a power-law decay with scaling exponent $\gamma$. 
In order to generate such a network ensemble, we treat the 
vertex degree   $k_i$ as a random integer variable governed by  the degree distribution
(\ref{eq:sf_degree_profile}). 

In order to generate a single random network with $N$ vertices, we used the following algorithm
for generating a vertex degree sequence consisting of $N$ vertex degrees $k_i$. We successively draw a random integer $k_i$ according to the degree distribution (\ref{eq:sf_degree_profile}). If the drawn vertex degree   satisfies $k_i \leq N-1$, we add it to our degree sequence. Otherwise, we discard it 
and draw another $k_i$ until we obtain one that fulfills this inequality. Thus, we use the upper 
cut-off $k_{\rm max}^{(1)} \equiv N-1$ for the vertex degree sequences. 
This procedure differs slightly from the one that we
used previously in \cite{Zhou-Lipowsky-2005}. In the latter study, we did not discard and redraw
the vertex degrees 
$k_i$ with $k_i > N-1$ but replaced them by $k_i = N-1$. This procedure has the 
disadvantage that  the actual degree distribution corresponding to the generated 
vertex degree sequence may develop a small peak at $k_i = N-1$ for sufficiently small
network size $N$.  The new algorithm used  here avoids this artefact. 
The choice for the upper cut-off $k_{\rm max}^{(1)} = N-1$
is rather natural since this is the maximal vertex degree for  a  network without 
multiple edges and self-connections. 
Note that we do not use an upper cutoff $k_{\rm max}^{(1)} \sim N^{1/2}$ as in
 Refs. \cite{Catanzaro-etal-2005,Castellano-PastorSatorras-2006}. 

For analytical estimates, it will be useful to define a second upper cut-off, 
 $k_{\rm max}^{(2)} \equiv k_0 N^{{1 \over \gamma-1}}$, 
for the vertex degree. The latter cut-off was
 introduced in  \cite{Cohen-etal-2000} and rederived by different arguments in 
\cite{Zhou-Lipowsky-2005}, see Appendix A. 
The effective   upper cut-off  for the 
 generated vertex degree sequence is then given by 
\be 
  k_{\rm max}  \equiv  \min\bigl(k_{\rm max}^{(1)}, k_{\rm max}^{(2)} \bigr) =  
  k_0 N^{{1 \over \gamma-1}}  
  \label{eq:k_max_true} 
\ee
where the  last equality applies to  $N > k_0^{(\gamma-1)/(\gamma-2)}$, and
the actual vertex degree distribution for  the generated degree sequence 
is   well approximated by 
\be 
  \label{eq:sf_degree_profile_2}
  P( k_i )={ k_i^{-\gamma} /   {\cal A}} 
\quad {\rm for } \quad k_0 \leq k_i \leq k_{\rm max} 
\ee
with the normalization constant ${\cal A}= \sum k_i^{-\gamma}$.
In the limit of large $k_{\rm max}$, the degree
distribution    (\ref{eq:sf_degree_profile_2}) is normalizable  provided  $\gamma > 1$,  and
the normalization constant is then given by 
\be 
  \label{eq:norm_A}
  {\cal A} \equiv \sum\limits_{k=k_0}^{k_{\rm max}} k^{-\gamma} \approx { k_0^{1-\gamma} - k_{\rm max}^{1-\gamma} \over \gamma-1} \ .
\ee
The mean vertex degree of the generated degree sequence is
\begin{equation}
  \label{eq:k_mean}
  \langle k \rangle = \sum\limits_{k=k_0}^{k_{\rm max}} k P( k ) \approx k_0 { (\gamma-1) \bigl( 1- (k_0/ k_{\rm max})^{\gamma-2} \bigr)
    \over (\gamma-2) \bigl( 1- (k_0 / k_{\rm max} )^{\gamma-1} \bigr)} \ .
\end{equation}
The mean vertex degree $\langle k \rangle $ attains a finite limit for large $k_{\rm max}$ provided 
$\gamma > 2$. It behaves as $ \langle k \rangle \approx   k_0 $ for large positive $\gamma$, corresponding to a random network
with uniform vertex degree $k = k_0$. On the other hand, the expression  (\ref{eq:k_mean})
also implies 
\begin{equation}
  \label{eq:k_mean_g2}
  \langle k \rangle \approx k_0 {\ln( k_{\rm max}/k_0 ) \over 1-(k_0/k_{\rm max}) } \hspace{0.5cm}{\rm for}\;\;\;\;
  \gamma=2 \ ,
\end{equation}
which diverges as  $\ln k_{\rm max}$ for large  $k_{\rm max}$.

\subsection{Multi-networks and simple-networks}
\label{sec:model_network}

After one degree sequence $\{ k_1,k_2,\ldots,k_N\}$ has been generated, we attach  $k_i$ 
half-edges to vertex $i$. Then we repeatedly crosslink two half-edges into a complete edge of the network, until
all  half-edges have been used up. In this way, an   initial network with the specified degree sequence is 
created which is then further randomized by performing a certain number of edge switching moves
\cite{Milo-etal-2003}.
We use two different ways  to implement the crosslinking process and the subsequent switching process
which lead to two different ensembles of networks,  multi-networks and simple-networks: 

{\em (A) Multi-networks:} During  the initial crosslinking process,   two half-edges are randomly chosen from the
  set of remaining half-edges and are then crosslinked into an edge of the network. 
  During the subsequent randomization or edge switching process, 
  two edges $\langle i j \rangle$ and $\langle k l \rangle$ are randomly chosen from the set of all edges of the network, and 
  they are replaced by two new edges $\langle ik \rangle$ and $\langle jl \rangle$. A network that 
  has been generated in this way will, in general, contain both 
  self-connections and multiple edges \cite{Zhou-2002}. 
  
{\em (B) Simple-networks:} During the initial crosslinking process, we again  randomly choose two half-edges from the set of remaining half-edges but crosslink them only into a tentative edge. This tentative edge 
is only kept if it provides a connection between two different vertices that have not been connected
before. Otherwise, the tentative edge is discarded. In this way, we discard all tentative
edges that represent self-connections or multiple edges. 
  During the subsequent randomization or edge switching process, we proceed in an 
  analogous manner: (i) we 
  randomly choose two edges $\langle i j \rangle$ and $\langle k l \rangle$ of the network; 
  (ii) we create two new
  tentative edges $\langle ik \rangle$ and $\langle jl \rangle$; and (iii) we keep these tentative
  edges only  if they do not represent self-connections or  multiple edges. Otherwise, 
 we reject this move and keep the old edges $\langle i j \rangle$ and $\langle k l \rangle$. 
 The networks generated in this way are simple in the sense that they contain no 
 self-connections and no multiple edges. 
 
 In principle, the network graphs generated by these two procedures could  have several disconnected components. In such a situation, the activity pattern would consist of  several  
subpatterns that are disconnected and, thus, evolve independently of each other. 
 The number and size of these components depends primarily on the scaling exponent $\gamma$ and the minimal vertex degree $k_0$. 
 In the following, we will discuss  ensembles of networks with  $2 < \gamma < 3$ and $k_0 \geq 5$ . For these parameter values,   all multi- and simple-networks that we generated  by the procedures (A) and (B) were found to consist of  only a single component. Each of these networks is then characterized by 
a single activity pattern in which all $N$ vertices participate. 
  
So far, we have not distinguished between a network and its most intuitive representation,  
 the corresponding  graph  with vertices $i$ and (undirected) edges $\langle ij \rangle$. 
 Another general representation   is provided by the adjacency matrix  $\mathbf I$  
 of the network. Each element $I_{ij}$ of this $N \times N$ matrix counts the number of  
 edges, $m(i,j)$, between vertex $i$ and vertex $j$, i.e., 
  \be
 I_{ij} \equiv m(i,j) \geq 0 . 
 \ee
 For an undirected graph  as considered here,  the adjacency matrix is symmetric 
 and $I_{ji} = I_{ij}$. Furthermore, a simple-network
 without self-connections and multiple edges is characterized by $I_{ii} = 0$ and 
  $0 \leq I_{ij} \leq 1$ for all vertex pairs $(i,j)$. 

\subsection{Local majority-rule dynamics}
\label{sec:model_majority_rule}

We now place a  binary or Ising-like spin 
$\sigma_i = \pm 1$ on each vertex $i$ of the network. Alternatively, we can use the 
activity variable $\sigma_i^\prime \equiv  (\sigma_i +1)/2$ which assumes the values 
$\sigma_i^\prime = 1$ and 0 corresponding to an active and inactive vertex $i$, 
respectively. Thus, the spin 
 configuration  $\{ \sigma (t) \} \equiv \{ \sigma_1(t), \sigma_2(t), \dots, \sigma_N(t) \}$ 
represents the activity pattern on the network  at time $t$. 

For the dynamics considered here, the 
time evolution of the spin or activity pattern is  governed by the local time-dependent fields 
\be
h_i (t)  \equiv {\sum_{j \neq i}}  I_{i  j} \sigma_j +  2 I_{i i} \sigma_i \, .
\label{localfield}
\ee
Because of the adjacency matrix  $\mathbf I$, the sum contains contributions from 
all vertices $j$ that are connected to vertex $i$, and this contribution is weighted 
by the multiplicity $I_{ij} = m(i,j)$ which counts the number of edges between $i$ and $j$. 
Therefore, 
the sign of the local  field $h_i$ is positive and negative if the {\em weighted  majority}
of the nearest neighbor spins is positive or negative, respectively. For multi-networks, 
the nearest neighbors are weighted with the corresponding edge multiplicity. For simple-networks, 
all nearest neighbors have the same multiplicity equal to one, and the absence of self-connections implies  $I_{i i} = 0$ for all vertices $i$. 

In the {\em parallel or synchronous} version of the local majority-rule dynamics, the discrete time 
$t$ is increased by $\Delta t \equiv  1$,  and the spin or activity pattern at  time $t$ is updated  simultaneously  on all vertices $i$ using the rule
\begin{eqnarray}
  \sigma_i(t+1) & \equiv & +1 \hspace{0.5cm}{\rm if} \;\; \;\; h_i(t) >0 \ , \nonumber \\
  & \equiv & -1 \hspace{0.5cm}{\rm if} \;\; \;\; h_i(t) < 0 \ .   \label{eq:parallel}
 \label{MajorityRuleParallel}
\end{eqnarray}
In the {\em random sequential or asynchronous}  version of the local majority-rule dynamics, time $t$ is 
increased by  $\Delta t \equiv 1/N$, a vertex $i$  is randomly chosen, and the spin value $\sigma_i$ on 
this vertex is  updated according to
\begin{eqnarray}
  \sigma_i(t+ 1/N) & \equiv & +1 \hspace{0.5cm}{\rm if} \;\; \;\; h_i(t) >0 \ , \nonumber \\
  & \equiv & -1 \hspace{0.5cm}{\rm if} \;\; \;\; h_i(t) < 0 \ .   \label{eq:sequential}
\label{MajorityRuleSequential}
\end{eqnarray}
while the spin values for all the other vertices remain unchanged during this update. In this way, 
the update of the whole network from time $t$ to time $t+1$ is divided up  into  $N$ 
successive substeps.  

If the local field $h_i(t)=0$ corresponding to an equal number of up and down spins on the 
nearest neigbor vertices, we choose $\sigma_i(t+\Delta t)=+1$ or $-1$ with equal probability
both  for the parallel update described by  (\ref{eq:parallel}) and for the random sequential 
update as given by (\ref{eq:sequential}). These local majority rules,  which are equivalent to Glauber dynamics \cite{Glauber-1963,BarYam-Epstein-2004} at  zero temperature,  have two fixed points corresponding to the 
two completely ordered patterns with $\{ \sigma (t) \} = \{ \sigma^{(-)} \} \equiv \{ - 1, -1, \dots, -1 \}$
and  $\{ \sigma (t) \} = \{ \sigma^{(+)} \} \equiv \{ +1, +1, \dots, +1 \}$

\subsection{Average properties of activity patterns}
\label{sec:average_properties}

In general, the
analysis of activity patterns on scale-free networks involves several types of 
averages. First, we consider an ensemble of networks which is characterized by the 
degree distribution $P(k)$. Second, 
we are interested in typical trajectories for the time evolution of the spin or activity pattern 
and, thus, consider an average over an ensemble of different initial patterns. 
Third, in order to characterize the global behavior of the patterns, we  perform  spatial 
averages, i.e.,   averages over the vertices $i$. 

The dynamics of the spin located on vertex $i$  is governed by the local field $h_i(t)$ as defined in (\ref{localfield}) which depends on the $k_i$  neighboring spins. 
Therefore, it will be useful to  divide the spatial average over all vertices up into 
averages over those vertices that have the same vertex degree $k$. The simplest 
average property is  the expectation value $\langle \sigma_i (t) \rangle$
which may be divided up according to 
\be
\langle \sigma_i (t) \rangle = \sum_k P(k)  \langle \sigma_i (t) \rangle_k
\equiv \sum_k P(k) [ 2 q_k(t) - 1]
\label{Defsmallqk(t)} 
\ee
where $ \langle \sigma_i (t) \rangle_k$ represents the average over all vertices with 
degree $k$ and $q_k(t)$ is the probability to  find a $k$--vertex in the spin-up state. 

The expectation value  $\langle \sigma_i (t) \rangle$ may also be viewed as 
an overlap function. In general, the  overlap of the actual pattern $ \{\sigma(t)\}$
with  an arbitrary   reference pattern  $\{{\hat \sigma} \}$ is given by 
 \be
 \Lambda ( \{\sigma(t)\}, \{{\hat \sigma} \}) \equiv \frac{1}{N} \sum_i \,  \sigma_i(t) \, {\hat \sigma}_i
 . 
 \ee
Thus, the expectation value $ \langle \sigma_i (t) \rangle$ measures the overlap
of the actual pattern with the completely ordered pattern 
$\{ \sigma^{(+)} \} = \{ +1, +1, \dots, +1 \}$
and 
\be
  \label{eq:Lambda}
\langle \sigma_i (t) \rangle = \Lambda  ( \{\sigma(t)\}, \{ \sigma^{(+)} \} ) 
\equiv \Lambda (t) . 
\ee

The average local field $\langle h_i (t) \rangle_k$ acting on a $k$--vertex depends on the 
expectation value $ \langle \sigma_i (t) \rangle_{k{\rm -ne}}$ where the subscript 
$k$--ne  indicates an average over all $k$ neighbors of the $k$--vertex. The latter 
expectation value can be expressed in terms of the $nn$--spin up probability $Q_k(t)$ that 
a randomly chosen neighbor of a $k$--vertex is in the spin-up state. One then 
has
\be
\langle h_i (t) \rangle_k = k \langle \sigma_i (t) \rangle_{k {\rm -ne}} 
\equiv k [ 2 Q_k(t) - 1] . 
\label{DefQk(t)}
\ee

The   probabilities $q_k (t)$ and $Q_k(t)$ are not independent but related via 
\be
Q_k(t) = \sum_{k^\prime}   P({k^\prime}|k) \, q_{k^\prime}
\label{Qkt}
\ee
with the conditional probability  $P({k^\prime}|k)$ that a  randomly chosen neighbor of 
a $k$--vertex is a ${k^\prime}$--vertex. In the following, the quantity $Q_k$ will be called 
the $nn$--spin up probability where the prefix `nn' stands for `nearest neighbour'. 

In the multi-networks as defined in Sect.  \ref{sec:model_network}
above, there are no vertex degree correlations. Following
a randomly chosen edge of a vertex with degree $k$, one will arrive at a vertex with degree
$k^\prime$. In such  uncorrelated random networks, the conditional probability $P(k^\prime | k)$ is independent of the vertex degree $k$, and one has
\begin{equation}
  \label{eq:Pk_conditional}
  P(k^\prime | k) = { k^\prime P(k^\prime)  /  \langle k \rangle } \ .
\end{equation}
In such a situation, the relation as given by  (\ref{Qkt})  simplifies and becomes
\be
Q_k(t) = \sum_{k}   \frac{ k P(k) }{ \langle k \rangle } \, q_{k} (t) 
\equiv Q(t)  .
\label{DefQ(t)} 
\ee
Therefore, for networks without vertex degree correlations,  the $nn$--spin up probability $Q_k(t) $
 is independent of $k$ and $Q_k(t)= Q(t)$ \cite{Zhou-Lipowsky-2005}.
The latter probability  satisfies the relation 
\be 
  \label{eq:Q}
2  Q(t) - 1 = \sum_{k}   \frac{ k P(k) }{ \langle k \rangle }  \langle \sigma_i (t) \rangle_k      \frac{ 1 }{ \langle k \rangle } \langle k_i \sigma_i \rangle  
\ee
where we used
the identity $ 2 q_k(t) - 1 = \langle \sigma_i (t) \rangle_k$. Thus, the $k$--independent
probability  $Q(t)$ is 
directly related to the weighted expectation value $ \langle k_i \sigma_i \rangle  $ which 
puts more weight on vertices with larger degree $k$. 

During the generation of a simple-network, 
compare Sect.  \ref{sec:model_network}, 
the elimination of self-connections and multiple edges
leads to correlations in the vertex degrees of  neighboring vertices. Such correlations
become significant for scale-free networks with $\gamma < 3$ 
\cite{Maslov-Sneppen-2002,Maslov-etal-2002,Boguna-etal-2004,Catanzaro-etal-2005}. 
Therefore,   the relation (\ref{eq:Pk_conditional})
represents an approximation for scale-free
random simple-networks with $\gamma <3$. In these latter
networks, the average vertex degree of the neighbors of a $k$--vertex,  $\langle k^\prime \rangle_{k{\rm -ne}}= \sum_{k^\prime} k^\prime P(k^\prime | k)$, is a
decreasing function of  $k$.

\section{Mean-field analysis}
\label{sec:analytic}

In the mean-field analysis, we use two simplifications. First, we assume that there are no 
correlations between the vertex degrees. Thus, we use the relationship 
  $P(k^\prime | k) = { k^\prime P(k^\prime)  /  \langle k \rangle }$ as in (\ref{eq:Pk_conditional}). 
  This implies that the $nn$--spin up probability $Q_k(t)$  is  
  independent of $k$ and identical to $Q(t) = \half [ 1 + \langle k_i \sigma_i \rangle  /  \langle k \rangle ]$.  
 As previously mentioned, this identity is  valid for the  multi-networks but represents
 an approximation for the simple-networks. 
 
 Second, we express  the probability $\cP(h_i(t) > 0)$  for the local field $h_i(t)$ 
 to be positive   in terms of  the probability $Q(t)$. In order to do so, 
 we sum over all  spin configurations of the neighboring vertices that 
 correspond  to a majority of up spins.  At this point, we assume that multi-edges are rare and 
 that   all $k_i$ neighboring spins are distinct. This assumption applies  to the simple-networks
 but represents an approximation for the multi-networks. Using this assumption, we obtain 
 for all vertices $i$ with vertex degree $k_i$ that
 \be
 \cP(h_i(t)  > 0) = \sum_{m = m_1}^{k}  B(k,m) \,  [Q(t)]^m [ 1 - Q(t) ]^{k-m} 
 \quad {\rm with} \quad k \equiv k_i
 \label{ProbPositive}
 \ee 
 where the summation over $m$ starts with 
 \be
 \begin{array}{lll@{\qquad{\rm for}\quad}l}
m_1 & \equiv & k_i/2  & {\rm   even \, \, } k_i \\ \nonumber
          &	\equiv & (k_i+1)/2       & {\rm   odd \, \, } k_i  \, , 
\end{array}  
 \ee
 and 
 \be
B(k,m) \equiv  \frac{k!}{ m! ( k - m)! } \equiv \Bin
\label{BinomialCoefficients}
\ee
 denotes the binomial coefficient. 
 
 If we now update the spin $\sigma_i(t)$ using the local majority rule as introduced above, we obtain $\sigma_i (t + \Delta t) = + 1$ with probability 
 \be
\cP_i^{(+)}(t + \Delta t) = \cP(h_i (t) > 0) + \half \cP(h_i(t) = 0)
\label{ProbUpdate}
 \ee
 and $\sigma_i (t + \Delta t) = - 1$ with probability $1 - \cP_i^{(+)}(t)$.  
If the vertex degree $k_i$ is odd, one cannot have the same number of neighboring up and 
down spins which implies that  the probability $\cP(h_i(t) = 0)$  vanishes
for all $t$.  If the vertex degree  $k_i $ is even, we obtain 
\be
\cP(h_i(t) = 0) = B(k, k/2) \, [Q(t)]^{k/2} \, [1 - Q(t)]^{k/2}  
\quad {\rm with} \quad k \equiv k_i
\label{ProbZero}
\ee
where $k_i$ distinct neighboring spins have been assumed as before.

\subsection{Parallel or synchronous updating}
\label{sec:analytics_parallel}

First, we apply the mean-field analysis just described to parallel or synchronous updating 
with $\Delta t = 1$ as defined by  (\ref{eq:parallel}). 
Inspection of the relations (\ref{ProbPositive}) and (\ref{ProbZero}) for the probabilities 
$ \cP(h_i(t)  > 0) $ and $\cP(h_i(t) = 0)$ shows that, within mean-field theory,  these probabilities 
are identical for all vertices with the same vertex degree $k_i = k$. Since these two probabilities
determine  the probability $\cP_i^{(+)}(t + \Delta t) $ via (\ref{ProbUpdate}), the latter probability 
is also identical for all vertices $i$ with the same $k_i = k$. Thus, mean field theory implies that 
 \be
\cP_i^{(+)}(t + \Delta t) = q_k(t + \Delta t)
\quad {\rm with } \quad 
k \equiv k_i  
\ee
where  the up spin probability $q_k$ has been defined in (\ref{Defsmallqk(t)}). After rearranging the summation over $m$, one   obtains the iteration equation
\be 
  \label{eq:q_k_t}
  q_k (t+ \Delta t) = {\sum_m}^\prime (1- \frac{1}{2} \delta_{m,k/2})
  \, B(k,m) \,
  Q^{m}(t) \bigl( 1-Q(t) \bigr)^{k-m}
\ee
where  the  prime  indicates that the summation now runs over
all integer $m$ with $k/2 \leq m \leq k$, and  $\delta$ is the Kronecker symbol. 
The iteration equation (\ref{eq:q_k_t}) is valid both for even and for odd values of $k$. 

If we insert the spin up probability $ q_k (t+ \Delta t) $ as given by (\ref{eq:q_k_t}) 
 with $\Delta t = 1$  into the relation 
(\ref{DefQ(t)}) for time $t+1$, we obtain the  evolution equation 
\be 
  \label{eq:Qt_evolution}
  Q(t+1) = \Psi \bigl(Q(t)\bigr)  
\ee
for the $nn$--spin up probability  $Q(t)$ where the evolution function $\Psi(Q)$ is defined by 
\begin{equation}
  \label{eq:Psi_Q}
  \Psi  (Q)  \equiv  \sum_k {\sum_m}^\prime
  (1- \frac{1}{2} \delta_{m,k/2})   \, k \,  P(k)   B(k,m) \,
  Q^m ( 1-Q)^{k-m} / \langle k \rangle  \ .
\end{equation}

The evolution equation~(\ref{eq:Qt_evolution}) has two stable
fixed points at $Q=0$ and  $Q = 1$, and an unstable one at $Q = 1/2$.
The fixed points with $Q=0$ and $Q = 1$ correspond  to the
all-spin-down and all-spin-up  pattern, respectively.
The unstable fixed point with $Q=1/2$ represents the phase boundary between
these two ordered patterns; the corresponding boundary patterns are
characterized
by probabilities ${\hat q_{k}}$ which satisfy
\begin{equation}
  \label{eq:PhaseBoundary}
  \sum_k \,  k \, P(k) \, {\hat q_{k}} ={\langle k \rangle/2}   \ .
\end{equation}

\begin{figure}
  \begin{center}
    \begin{minipage}{0.7\linewidth}
      \includegraphics[width=10cm]{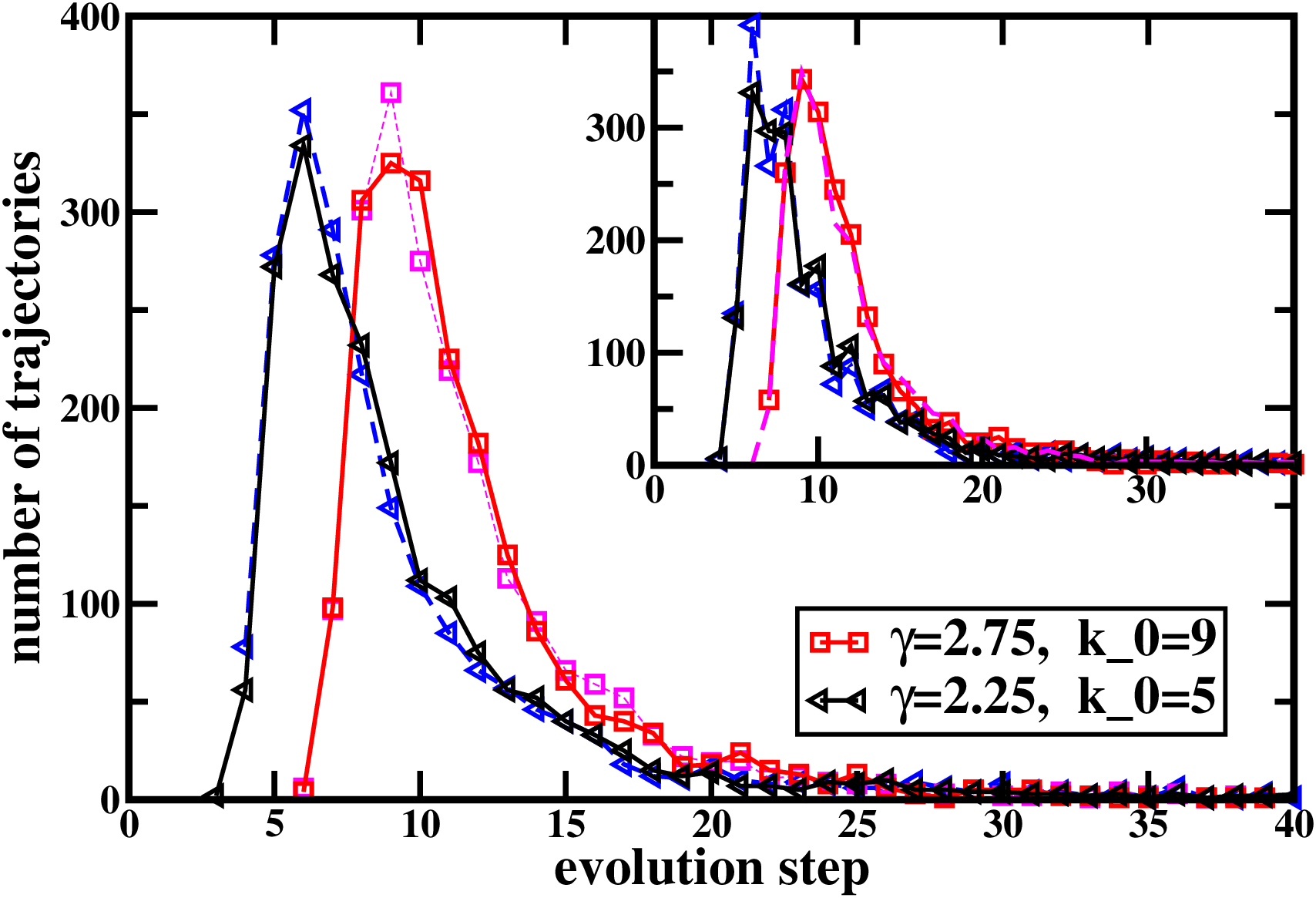}
    \end{minipage}
  \end{center}
  \caption{
    \label{fig:time_profile} 
    Histogram of decay times $t_d$ for parallel local majority-rule dynamics on a 
    uncorrelated random multi-network. The decay time is measured in units of iteration steps. 
    The four sets of data   are obtained for two random 
    scale-free networks with $\gamma=2.25$,
    $k_0=5$ (triangles) and two random scale-free networks with 
    $\gamma=2.75$, $k_0=9$ (squares). All four networks have the same vertex number 
    $N=2^{18}$ and almost the same mean vertex
    degree $\langle k \rangle \simeq 20$. For each data set, $2,000$ individual trajectories
    with the initial condition of $\Lambda=0$ and $Q=1/2$ are simulated.
    Each trajectory is tracked until the shifted $nn$--spin up probability $|Q(t)-1/2|$ exceeds $1/4$
    for the first time (the main figure) or  the average spin value $|\Lambda(t)|$ exceeds
    $1/2$ for the first time (the inset).
  }
\end{figure}

\begin{figure}
  \begin{center}
    \begin{minipage}{0.7\linewidth}
      \includegraphics[width=10cm]{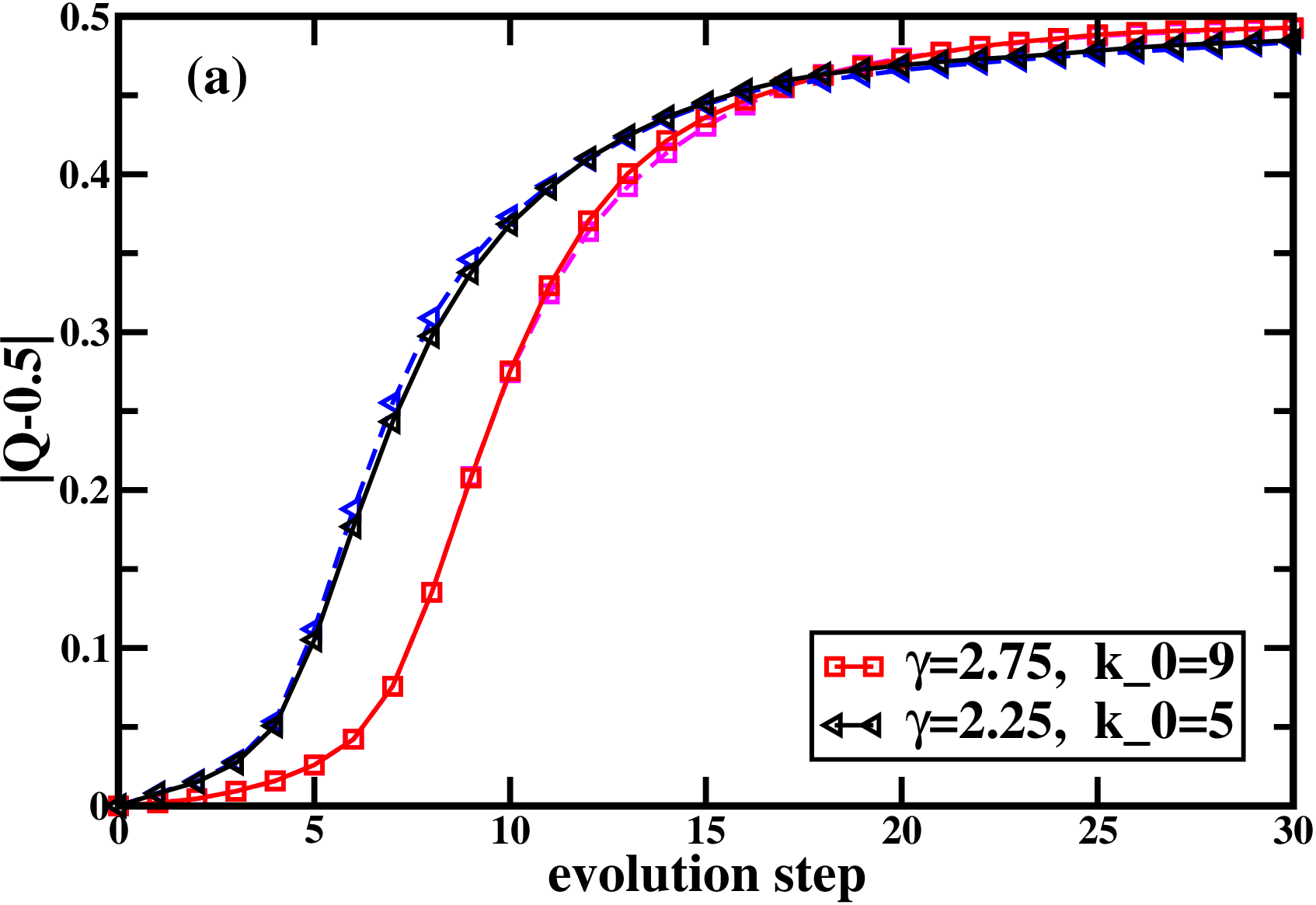}
  \vskip 0.5cm
      \includegraphics[width=10cm]{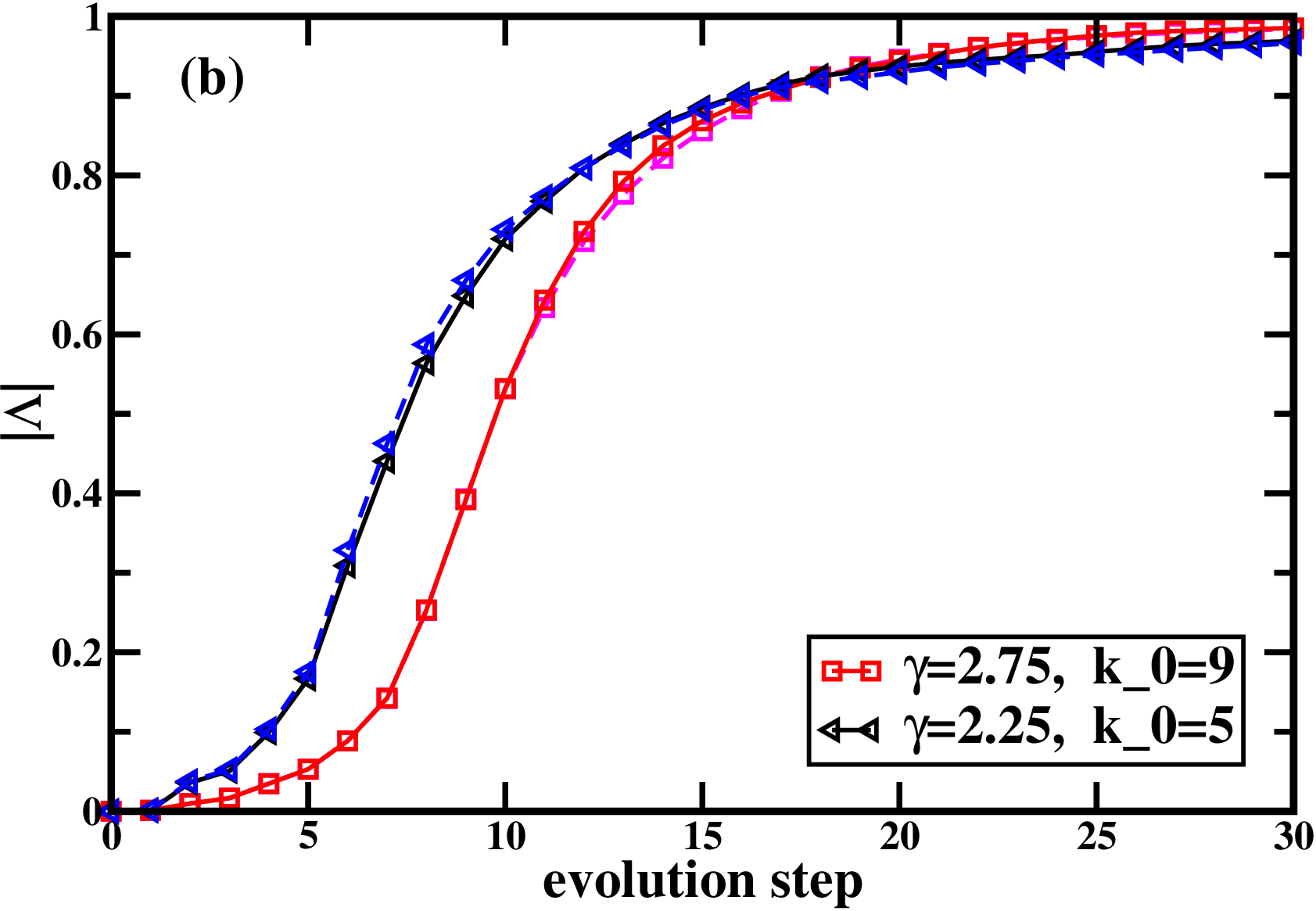}
    \end{minipage}
  \end{center}
  \caption{
    Time evolution of (A) the  $nn$--spin up probability  $Q(t)$   and (B)
     the average spin value $\Lambda(t)$   for  parallel local majority-rule
    dynamics on an uncorrelated random multi-network of size $N=2^{18}$ and mean vertex degree
    $\langle k \rangle \simeq  20$. Time is measured in units of iteration steps. 
    The four sets of data   correspond to the same   four scale-free random networks as in  
    Fig.~\ref{fig:time_profile}. 
  }
 \label{fig:QA_change}
\end{figure}

Since  $Q=1/2$ represents an unstable fixed point for the time evolution of the 
$nn$--spin up probability $Q(t)$, 
a natural question to ask is the typical time   needed for the
system to escape from $Q=1/2$. Figure~\ref{fig:time_profile} displays the 
histogram of the   decay times $t_d$ needed for the system to evolve from $Q(t_0)=1/2$
at time $t_0$ to $|Q(t_0+t_d) -1/2| > 1/4$ at time $t_0+t_d$. For the four scale-free random networks
with mean vertex degree $\langle k \rangle \simeq  20$, we find that the typical decay time is positively
correlated with the  scaling exponent $\gamma$. The most probable decay
time, for example,  is $t_{d,mp} =6$ for $\gamma=2.25$ but $t_{d, mp} =9$ for  $\gamma=2.75$. 
Figure~\ref{fig:QA_change}
shows the mean value of $Q(t)$ and $\Lambda(t)$ as a function of time $t$ for the same
four different random networks. As the scaling exponent $\gamma$ becomes smaller, the
system approaches the completely ordered patterns more quickly.

To understand this difference in decay times, we now estimate the mean decay time according
to our mean-field theory. For this purpose, we define the  order
parameter  
\begin{equation}
  \label{eq:y}
  y \equiv Q-1/2 \ .
\end{equation}
The order parameter $y$ vanishes for all the boundary patterns satisfying
Eq.~(\ref{eq:PhaseBoundary}).
In the vicinity of $y=0$, the evolution of $y$ is governed by the
following linearized evolution equation
\be 
  \label{eq:evolution_y_0}
  y(t+1)= \Psi^\prime(1/2) y (t) \ ,
\ee
where $\Psi^{\prime}(1/2) \equiv {\rm d} \Psi / {\rm d} Q |_{Q=1/2}$ is expressed as 
\begin{eqnarray}
  \Psi^\prime(1/2)&=& \sum\limits_{k} \sum\limits_{m> k/2}^{k} { k P(k) \over \langle k \rangle } B(k,m) 2^{1-k} (2 m - k)
  \nonumber \\
  &=& \sum\limits_{m=0}^\infty {(2 m+1)^2 P( 2 m +1 ) \over \langle k \rangle 2^{2 m} } B(2m,m)
  + \sum\limits_{m=1}^\infty{(2 m)^2 P(2 m) \over \langle k \rangle 2^{2 m}} B(2m,m) \ .
  \label{eq:psi_derivation_half}
\end{eqnarray}
The binomial coefficient in Eq.~(\ref{eq:psi_derivation_half}) 
can be estimated by Stirling's formula which leads to
\be 
  \label{eq:B_2m_m}
  B(2 m, m) = {(2 m) ! \over m! m!} \approx {2^{2 m} \over \sqrt{\pi}} m^{-1/2} \ .
\ee
If this relation is inserted into Eq.~(\ref{eq:psi_derivation_half}), we obtain
\be 
  \label{eq:Psi_derivation_half_scale}
  \Psi^{\prime}(1/2) \approx { \sqrt{2/ \pi} \over \langle k \rangle } \sum\limits_{k} 
  k^{{3\over 2}} P(k) \ .
\ee
 
In the limit of large network size $N$, Eq.~(\ref{eq:Psi_derivation_half_scale}) 
is well approximated by
\begin{equation}
  \label{eq:Psi_d_Q_app}
  \Psi^\prime (1/2) \approx \bigl( { 2 k_0 \over \pi} \bigr)^{1/2} { (\gamma -2 ) \over \bigl(
    {5 \over 2} - \gamma \bigr)}
  {(k_{\rm max} / k_0)^{(5-2 \gamma)/2} - 1 \over 1- (k_0 / k_{\rm max} )^{\gamma-2} } \ .
\end{equation}
In the limiting case of $\gamma=5/2$ and $\gamma=2$, the expressions for $\Psi^\prime(1/2)$ is, respectively,
\begin{eqnarray}
  \Psi^\prime(1/2) &\approx & \bigl({k_0 \over 2 \pi} \bigr)^{1/2} { \ln ( k_{\rm max} / k_0 ) \over 
    1-(k_0 / k_{\rm max})^{1/2} } \hspace{0.5cm}{\rm for}\;\;\;\; \gamma=5/2 \label{eq:dPhi_1} \\
  \Psi^\prime (1/2) & \approx & 2 \bigl({2 k_0 \over \pi} \bigr)^{1/2}
  {(k_{\rm max} / k_0)^{1/2} - 1 \over \ln ( k_{\rm max} / k_0 )} \hspace{0.5cm}{\rm for}\;\;\;\;
  \gamma=2 \ . \label{eq:dPhi_2} 
\end{eqnarray}
Finally, using the relation that $k_{\rm max}= k_0 N^{1/(\gamma-1)}$ as given by
Eq.~(\ref{eq:k_max_true}) for sufficiently large network size $N$, 
we obtain the following asymptotic
behavior for $\Psi^\prime (1/2)$
\begin{eqnarray}
  \Psi^\prime (1/2) &\approx & 2 \sqrt{ 2 k_0 / \pi} N^{1/2} / \ln (N) \hspace{0.5cm}  {\rm for} \;\;\;\;\gamma=2 \ , 
  \nonumber\\
  & \approx & \sqrt{ 2 k_0 / \pi} {\gamma-2 \over {5\over 2} - \gamma} N^{{5-2 \gamma \over 2 ( \gamma-1)}} \hspace{0.5cm}
  {\rm for} \;\;\;\; 2 < \gamma < 5/2 \ , 
  \nonumber  \\
  &\approx & {1 \over 3} \sqrt{2 k_0 / \pi} \ln (N) \hspace{0.5cm}  {\rm for} \;\;\;\; \gamma=5/2 \ , 
  \nonumber\\
  & \approx & \sqrt{2 k_0 / \pi} {\gamma-2 \over \gamma-{5\over 2}} \hspace{0.5cm} {\rm for} \;\;\;\;\gamma > 5/2 \ .
  \label{eq:dPsi_a}
\end{eqnarray}
From Eq.~(\ref{eq:dPsi_a}) we know that $\Psi^\prime(1/2)$ is independent of network size $N$ for
$\gamma >5/2$ while it increases with $N$ for $\gamma \leq 5/2$. Since the slope $\Psi^\prime(1/2)$ determines the
evolution of $y(t)$ for small $y$, we expect to see a qualitative change at $\gamma=5/2$.

Iterating  the linear evolution equation (\ref{eq:evolution_y_0}) $n$ times from an initial time $t_0$ up to a final time
$t_1$, one obtains the time difference
\begin{equation}
  \label{eq:time_difference}
  \Delta t_{01} \equiv  t_1 - t_0 \approx { \ln|y(t_1)| - \ln | y(t_0)| \over \ln \Psi^\prime (1/2)} 
\end{equation}
in the limit of small $y(t_0)$.

Now consider an initial state of the network at time $t=t_0$ that corresponds to a strongly
disordered pattern with order parameter $y(t_0)=\pm 1/(2 M) = \pm 1/ \langle k \rangle N$
where $M$ denotes the total number of edges with $2M = \sum_i k_i   $. 
We characterize the decay of 
this strongly disordered patterns by the decay time $t_d$, which is the time it takes to
reach a pattern with an order parameter $y_{*}$ that satisfies $|y_{*}|\geq 1/4$.
Using Eq.~(\ref{eq:time_difference}) with $t_1 \equiv t_d$ as well as the asymptotic 
expressions (\ref{eq:dPsi_a}) for $\Psi^\prime(1/2)$, the functional dependence
of the decay time 
$t_d$ on the network size $N$ is found to be \cite{Zhou-Lipowsky-2005}
\be
\begin{array}{lll@{\qquad{\rm for}\quad}l}
 t_d  & \sim  &   \ln (N)   &
                                                         \gamma > 5/2 \\[2.5ex]  \nonumber
          &	\sim  &   \ln (N) / \ln \ln (N)      & 
                                                           \gamma = 5/2 \\[2.5ex] \nonumber
          &	\sim  &   2 (\gamma - 1)/ (5 - 2\gamma)    & 2 < \gamma < 5/2    \\ 
\end{array}  
\label{eq:time_scale}
\ee
in the limit of large  $N$.
Equation (\ref{eq:time_scale}) predicts that, for  random scale-free networks
with  $2 < \gamma < 5/2$, strongly disordered patterns always escape from the
unstable fixed point after a  {\em finite} number of  iteration steps even in the limit of 
large $N$. In contrast,
for networks with $\gamma > 5/2$, the escape time diverges as $\ln(N)$. This latter behavior also
applies to Poisson networks and other types of networks with single-peaked vertex degree
distributions. The $\ln(N)$ scaling in response times was predicted 
for opinion spreading on social networks 
\cite{Krapivsky-Redner-2003,Aldana-Larralde-2004,Sood-Redner-2005}, it
was also observed in simulations of information spreading on small-world networks
\cite{Zhu-etal-2004}.

We will check these mean-field predictions in Sec.~\ref{sec:simulation_MRN} and
Sec.~\ref{sec:simulation_SRN} by extensive computer simulations.

\subsection{Random sequential or asynchronous updating}
\label{sec:analytics_sequential}

Next, we apply our mean-field analysis to random sequential or asynchronous updating. 
In this case, we  randomly choose single spins. Each single spin update 
corresponds to the time step $\Delta t = 1/N$ which becomes small for large $N$. It is now useful to consider the quantity $\Delta n(k,t)$
which represents the change in the number of up spins on all $k$--vertices during such a 
single spin update. The average value of this quanitity is given by 
\be
\langle  \Delta n(k,t) \rangle = P(k) \, [1 - q_k(t)] \,  q_k (t + \Delta t) - P(k) \, q_k(t) \, [ 1 - q_k(t + \Delta t)]
\, .
\label{Deltan(k,t)}
\ee
The total number of up  spins on all $k$--vertices then changes according to 
\be
N P(k) q_k (t + \Delta t)  = N P(k) q_k(t) + \langle  \Delta n(k,t) \rangle \, .
\ee
Using the expansion  $q_k (t + \Delta t) \approx q_k(t) + \Delta t \, \rd q_k(t) /\rd t $
for large $N$ or small $\Delta t = 1/N$
on the left hand side and  inserting the expression (\ref{Deltan(k,t)}) for $\Delta n(k,t) \rangle$
on the right hand side, we obtain
\be
 \frac{\rd q_k(t)}{\rd t} =  - q_k(t) + q_k (t + \Delta t) \, .
\ee
In addition, the up spin probability $q_k (t + \Delta t)$ is still given by  (\ref{eq:q_k_t}) which 
leads to the continous-time evolution equations 

\be
  \label{eq:q_k_change}
  {{\rm d} q_k(t) \over {\rm d} t} = - q_k(t) + {\sum_m}^\prime (1- \frac{1}{2} \delta_{m,k/2})
  \, B(k,m) \,
  [Q(t)]^{m} \bigl[ 1-Q(t) \bigr]^{k-m}  
\ee
 for the spin up probabilities $q_k(t)$ and
 \be 
  \label{eq:Qt_change}
  {  {\rm d} Q(t) \over {\rm d} t} = -Q(t) + \Psi\bigl( Q(t) \bigr)  
\ee
for the $nn$--spin up probability $Q(t)$ 
where $\Psi(Q)$ is still given by (\ref{eq:Psi_Q}). Equation (\ref{eq:Qt_change})  has again 
three fixed points, two stable ones at $Q=0$ and  $Q = 1$ as well as an unstable one at $Q=1/2$. 

As in the scheme with parallel updating, we define a new order parameter
$y = Q - 1/2$, compare (\ref{eq:y}). For an initial value $y(t_0) \sim 1/N$ that represents a 
strongly disordered initial pattern, 
the order parameter   $y(t)$ behaves as
\be 
  \label{eq:y_change}
  y(t_1) \approx  y(t_0) e^{( \Psi^\prime (1/2)-1 ) (t_1-t_0)}  
\ee
which implies the time difference  
\be
  \Delta t_{01}\equiv t_1-t_0 \approx {\ln y(t_1)-\ln y(t_0) \over \Psi^\prime (1/2)-1} \ .
 \label{DeltaTime2}
\ee
We again  characterize the decay of 
this strongly disordered patterns by the decay time $t_1 = t_d$, which is the time it takes to
reach a pattern with an order parameter $y_{*}$ that satisfies $|y_{*}|\geq 1/4$.
Using the initial value  $y(t_0) \sim 1/N$, we now obtain
from (\ref{DeltaTime2}) that  the typical decay time $t_d$  scales as
\be
\begin{array}{lll@{\qquad{\rm for}\quad}l}
 t_d  & \sim  &   \ln (N)   &
                                                         \gamma > 5/2 \\[2.5ex]  \nonumber
          &	\sim  &  N^0     & 
                                                           \gamma = 5/2 \\[2.5ex] \nonumber
          &	\sim  &  N^0    & 2 < \gamma < 5/2    \\ 
\end{array}  
 \label{eq:td_scale_2} 
\ee
in the limit of large network size $N$.  Equation~(\ref{eq:td_scale_2}) is qualitatively
very similar to the scaling relation  (\ref{eq:time_scale}) for the
parallel updating scheme. A transition in dynamic behavior is therefore also
predicted for random sequential local majority-rule dynamics.
For uncorrelated scale-free random networks with
$\gamma >5/2$, Eq.~(\ref{eq:td_scale_2}) again predicts a logarithmic scaling of the typical
decay time $t_d$ with network size $N$. When the scaling exponent $\gamma < 5/2$, 
on the other hand, the typical decay time is independent of $N$. The case $\gamma = 5/2$
is special since the mean-field analysis predicts that $t_d$ is independent of $N$ for 
random sequential updating but grows logarithmically with $N$ for parallel updating.

\section{Simulation results for  random multi-networks}
\label{sec:simulation_MRN}

As a complementary approach to our mean-field analysis in Sec.~\ref{sec:analytic}, in this and
the following section, we perform local majority-rule dynamics on individual random scale-free
networks. In this section, the substrate networks are generated by rule A of 
Sec.~\ref{sec:model_network}. These multi-networks are completely random and uncorrelated, 
and typically contain both self-connections and multiple edges.
 
After a network is constructed with given parameters $N$, $\gamma$, and $k_0$, a random initial spin
pattern $\{ \sigma_i(0) \}$ is assigned to vertices of the network. The system then evolves according to
Eq.~(\ref{eq:parallel}) for parallel updating and according to  Eq.~(\ref{eq:sequential})  for random
sequential updating. Each resulting  trajectory  is monitored and the values of 
$\Lambda(t)$ and $Q(t)$ as defined in Eq.~(\ref{eq:Lambda}) and Eq.~(\ref{eq:Q}), respectively, are
recorded at each time point $t$. The initial spin patterns are chosen such that the  average 
spin value $\Lambda $ and the  $nn$--spin up probability $Q$ have the initial values 
\be 
  \label{eq:zero_AQ}
  \Lambda(0)\equiv 0
  \quad {\rm and } \quad Q(0)\equiv{1 \over 2} \ ,
\label{StronglyDisorderedPattern}
\ee
i.e., these patterns are  disordered   both with respect to  the average spin value and with 
respect to the $nn$--spin up probability.   We track 
each  trajectory until  the two conditions $|\Lambda(t)| > 1/2$ and $|Q(t)-1/2| > 1/4$
are both satisfied. For each random network, $2,000$ such trajectories are simulated and 
analyzed in order to estimate  the median decay time.

\begin{figure}[bth]
  \includegraphics[width=10cm]{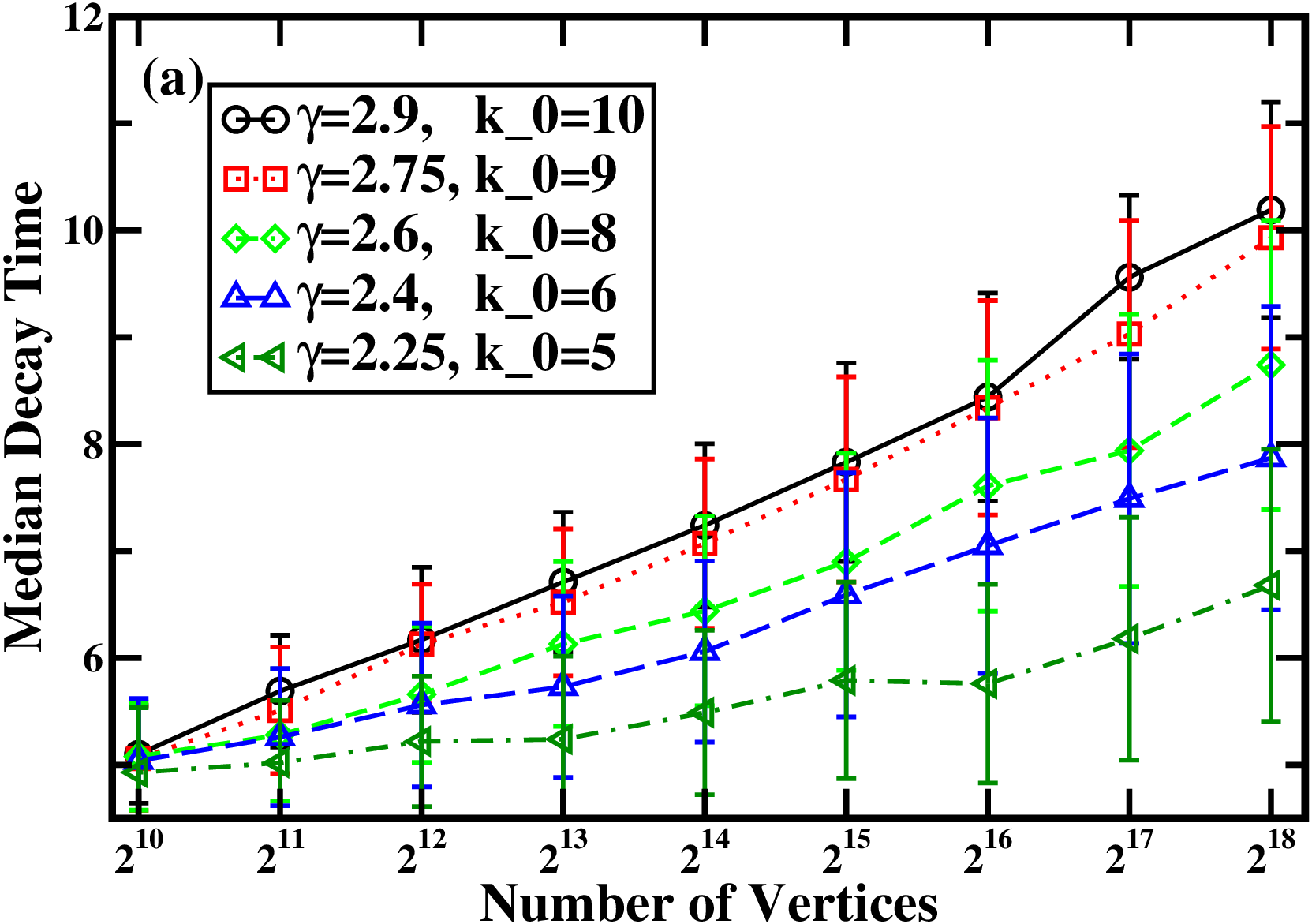}
  \vskip 0.5cm
  \includegraphics[width=10cm]{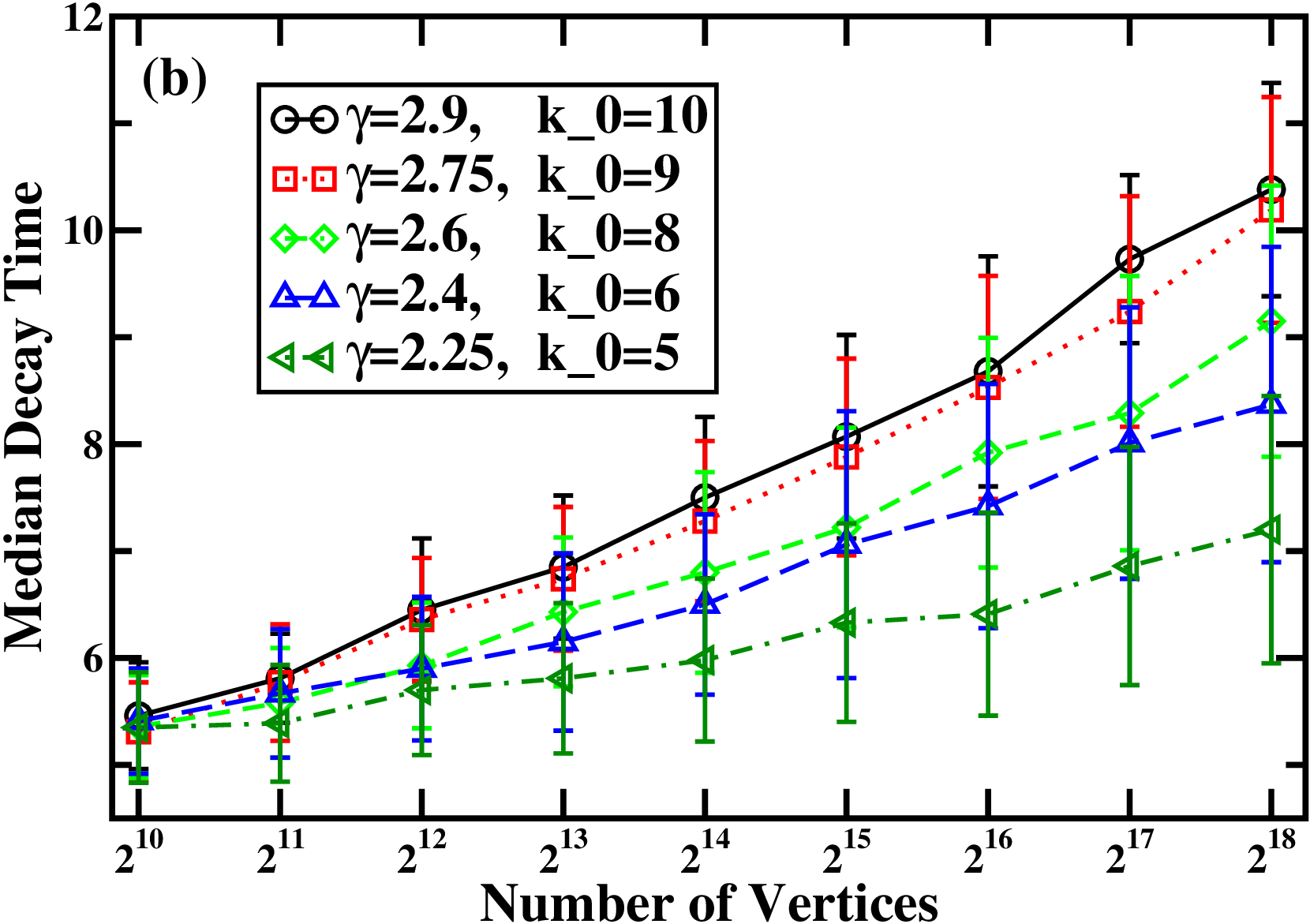}
  \caption{
    \label{fig:QA_MRN_p}
    Median decay time $t_{d,m}$ as a function of network size $N$ 
    for random multi-networks  and 
    parallel or synchronous  updating. Each data point is obtained by performing
    simulations on $100$  multi-networks.
      (a) Median decay time needed to reach a partially ordered pattern characterized by  $|Q(t)-1/2 >  
      1/4|$  for the
    first time; and (b) Median decay time needed to reach a partially ordered pattern with  
    $|\Lambda (t) > 1/2|$  for the first time.
  }
\end{figure}

\begin{figure}[bth]
  \includegraphics[width=10cm]{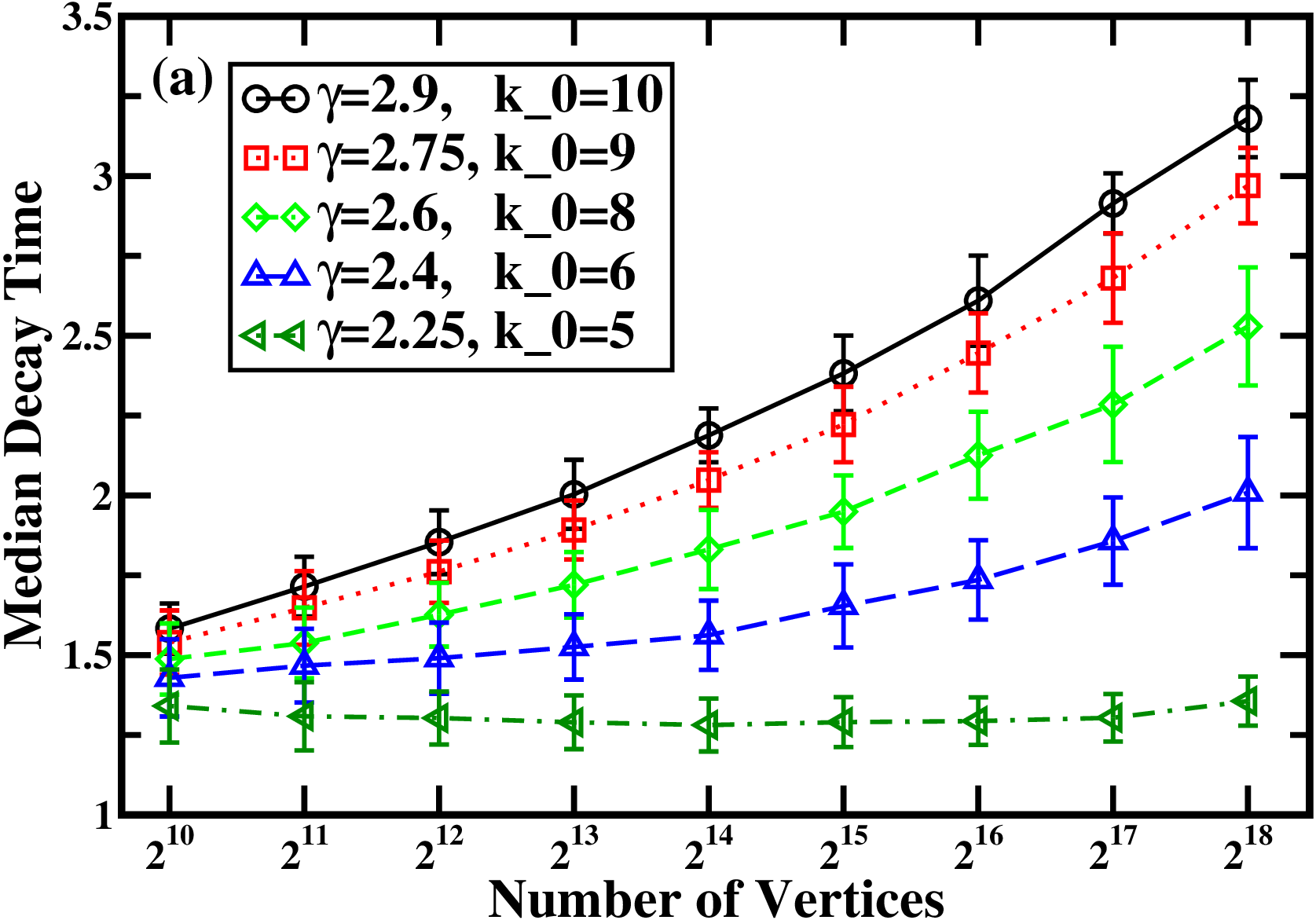}
  \vskip 0.5cm
  \includegraphics[width=10cm]{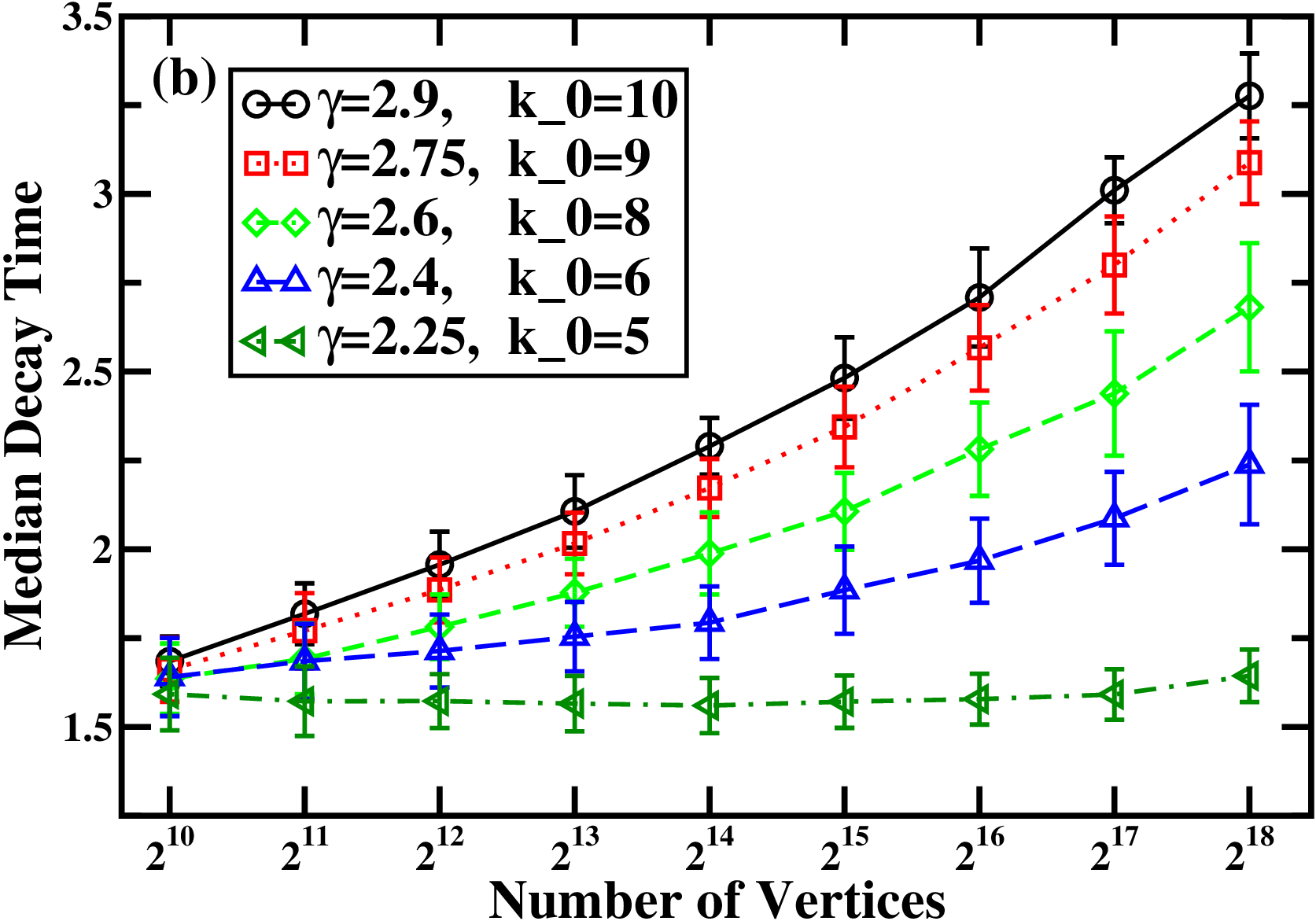}
  \caption{
    \label{fig:QA_MRN_s}
    Median decay time $t_{d,m}$ as a function of network size $N$ 
    for random multi-networks  and 
   random sequential or asynchronous updating. Each data point is obtained by performing
    simulations on the same set of
    $100$ random multi-networks as in Fig.~\ref{fig:QA_MRN_p}.
    (a) Median decay time needed for $|Q(t)-1/2|$ to exceed $1/4$ for the
    first time. (b) Median decay time needed for $|\Lambda(t)|$ to exceed $1/2$
    for the first time. 
  }
\end{figure}

In our previous study \cite{Zhou-Lipowsky-2005},  we used parallel or synchronous updating in order to generate the time evolution of the activity patterns. In the present study, we use both parallel updating and random sequential or asynchronous updating and check the robustness of 
 our  qualitative conclusions in Ref.~\cite{Zhou-Lipowsky-2005}. 
For parallel  updating, our simulational results are summarized in
Fig.~\ref{fig:QA_MRN_p},  for random sequential  in Fig.~\ref{fig:QA_MRN_s}. In these figures, 
we plot the median decay time, $t_{d,m}$, as a function of the network size or vertex number $N$. 
By definition, 
the median decay time  $t_{d,m}$ splits  the decay time histogram up into two 
parts of equal size, i.e., the probability to observe a decay time $t_d$ with $t_d < t_{d,m}$
is equal to the probability to observe one with  $t_d > t_{d,m}$. 
In Fig.  \ref{fig:QA_MRN_p}(a) and Fig.~\ref{fig:QA_MRN_s}(a), we display 
the median decay time needed for the system to reach a pattern characterized by 
$|Q(t)-Q(0)| \geq 1/4$ from an initial strongly disordered pattern, which satisfies  Eq.~(\ref{eq:zero_AQ}). Likewise,  in Fig.~\ref{fig:QA_MRN_p}(b)
and Fig.~\ref{fig:QA_MRN_s}(b), we show 
the median decay time for the system to attain a pattern characterized by  $|\Lambda(t)| \geq 1/2$ starting again from an 
initially disordered   pattern satisfying Eq.~(\ref{eq:zero_AQ}).

The data for   parallel updating of multi-networks as shown in Fig.~\ref{fig:QA_MRN_p}
do  not exhibit a clear distinction between networks with $\gamma > 5/2$ and those with 
$\gamma < 5/2$.  In contrast, the data for random sequential updating of multi-networks 
as shown in Fig.~\ref{fig:QA_MRN_s} exhibit  such a distinction, at least for $\gamma = 2.25$
and $\gamma = 2.6$. 
For  scaling exponent $\gamma \geq  2.6$,
the median decay times $t_{d,m}$ increases with network size $N$ according to $t_{d,m} \sim (\ln N)^\eta$, where the exponent
$\eta$ appears to be slightly larger than $1$. Our mean field theory predicts $\eta=1$. 
For  $\gamma = 2.25$, which is smaller than but not   close to  $\gamma = 5/2$, the median decay time $t_{d,m}$ is found to be independent of network size $N$, see Fig.~\ref{fig:QA_MRN_s}(a) and 
(b), 
in agreement with our mean-field prediction. However, for $\gamma=2.4$,  which is smaller than but close to 
$\gamma = 5/2$, inspection of  Fig.~\ref{fig:QA_MRN_s} indicates 
that the median decay time  increases slowly with network size $N$. This observation 
disagrees with  our mean field prediction.

\section{Simulation results for random simple-networks}
\label{sec:simulation_SRN}

The random multi-networks discussed in the previous Sec.~\ref{sec:simulation_MRN} 
contain
self-connections and multiple edges between the same pair of vertices. Such connections are
absent in many real networks. Thus, we will now consider  
the temporal evolution of activity patterns on  random simple-networks without self-connections and multiple edges. As previously mentioned, such networks exhibit some vertex degree correlations. 

\begin{figure}[bth]
 \includegraphics[width=10cm]{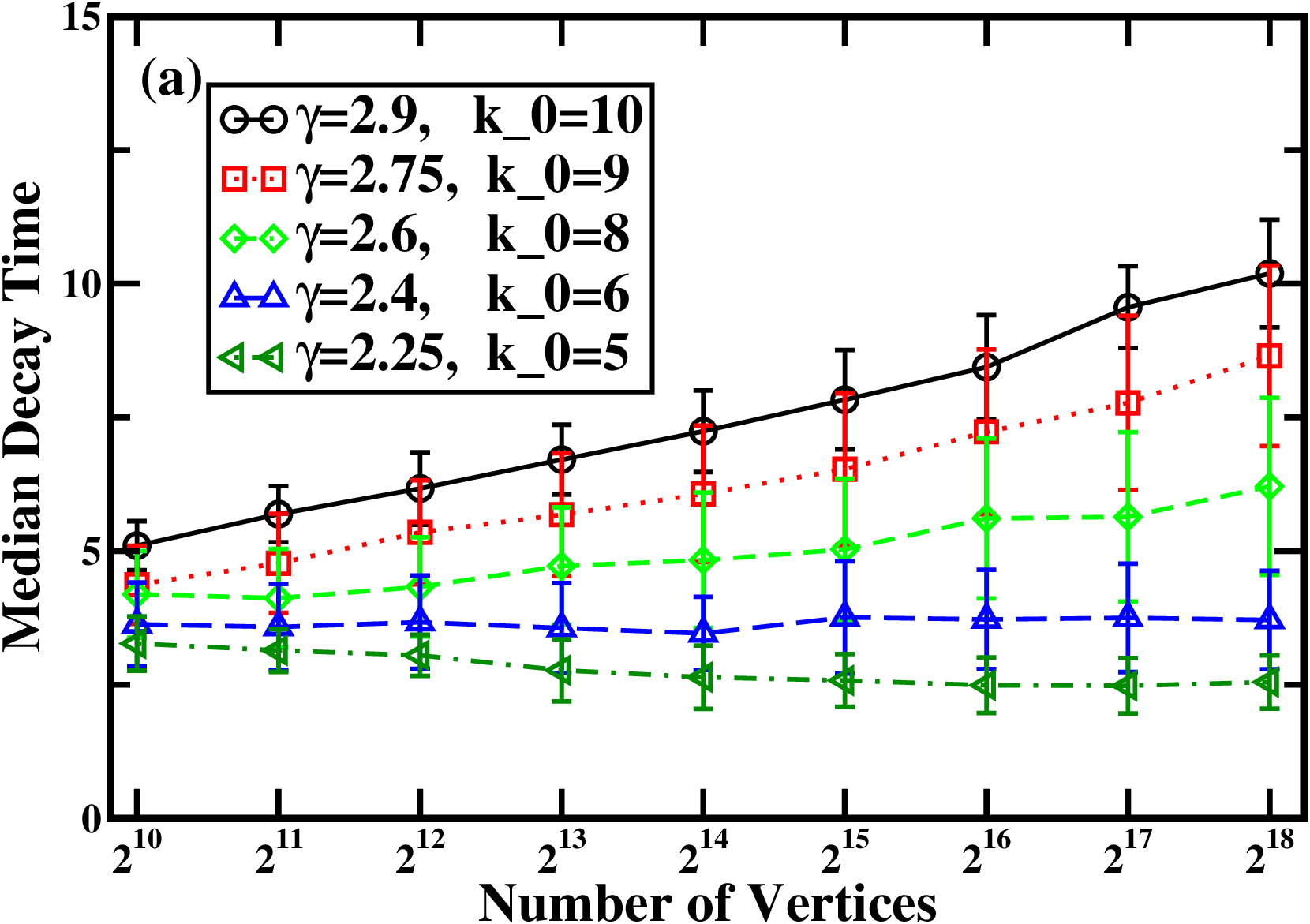}
  \vskip 0.5cm
  \includegraphics[width=10cm]{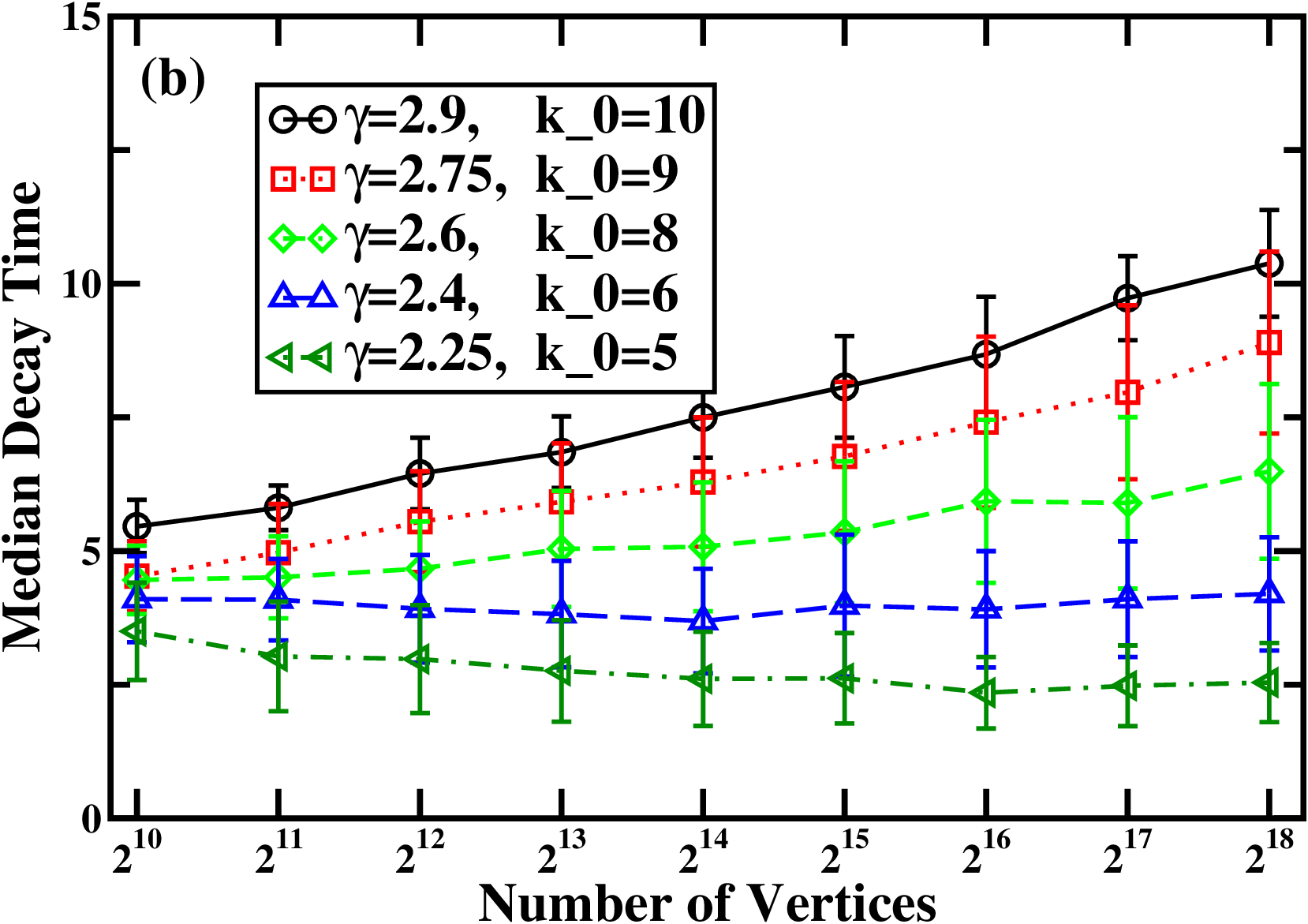}
  \caption{
    \label{fig:QA_SRN_p}
    Median decay time as a function of network size $N$ for random 
    simple-networks and 
    parallel or synchronous updating. Each data point is obtained by performing
    simulations for an ensemble of $100$  networks. Each simulation run starts from an initial  spin or activity pattern characterized by relation  (\ref{eq:zero_AQ}). 
    (a) Median decay time needed to reach a spin pattern for which   the $nn$--spin up probability $Q(t)$ satisfies $|Q(t)-1/2| > 1/4$ for the
    first time; and (b) Median decay time needed to reach a spin pattern for which the average spin value $\Lambda(t)$  exceeds $1/2$
    for the first time.
  }
\end{figure}

\begin{figure}[bth]
   \includegraphics[width=10cm]{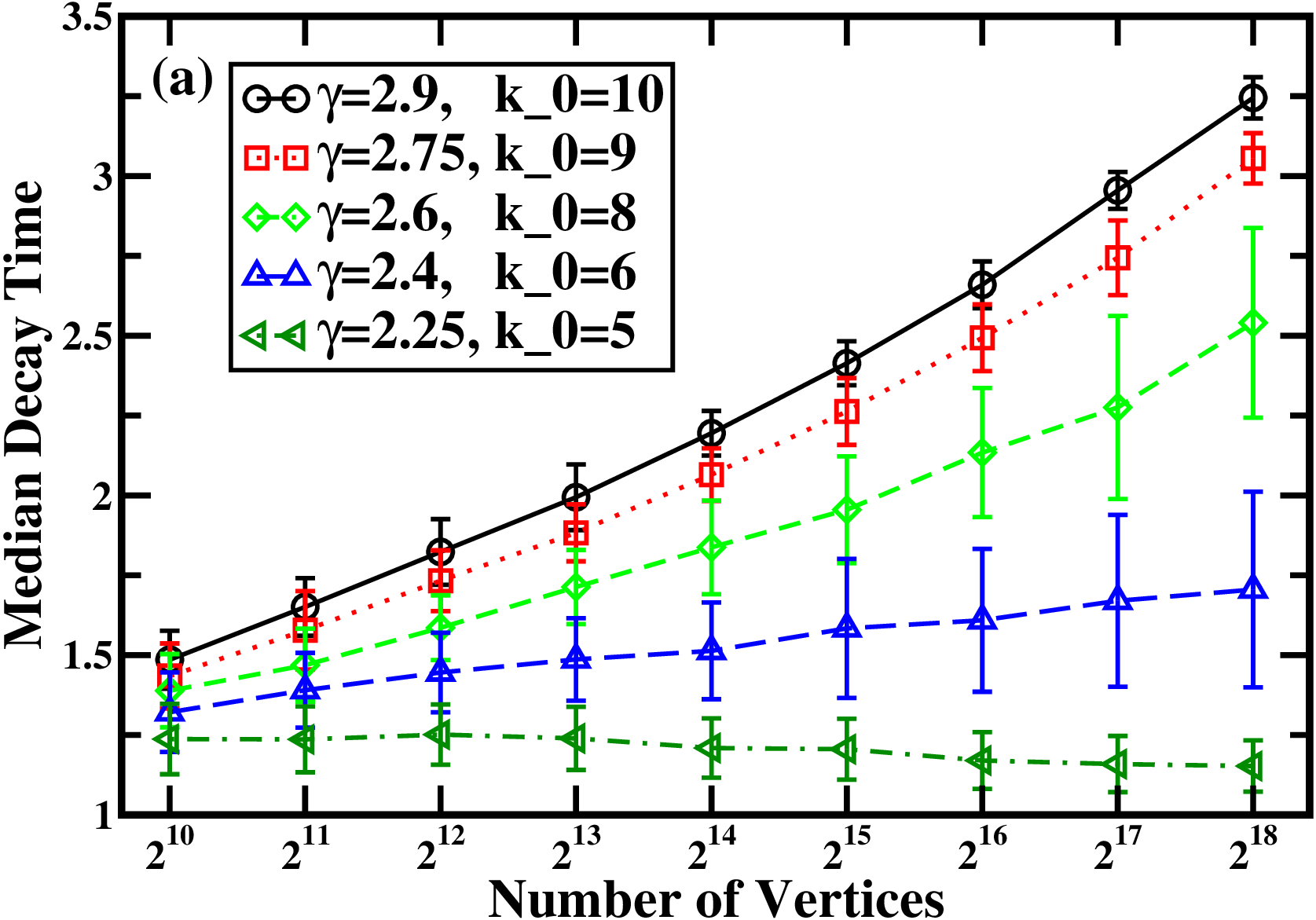}
  \vskip 0.5cm
   \includegraphics[width=10cm]{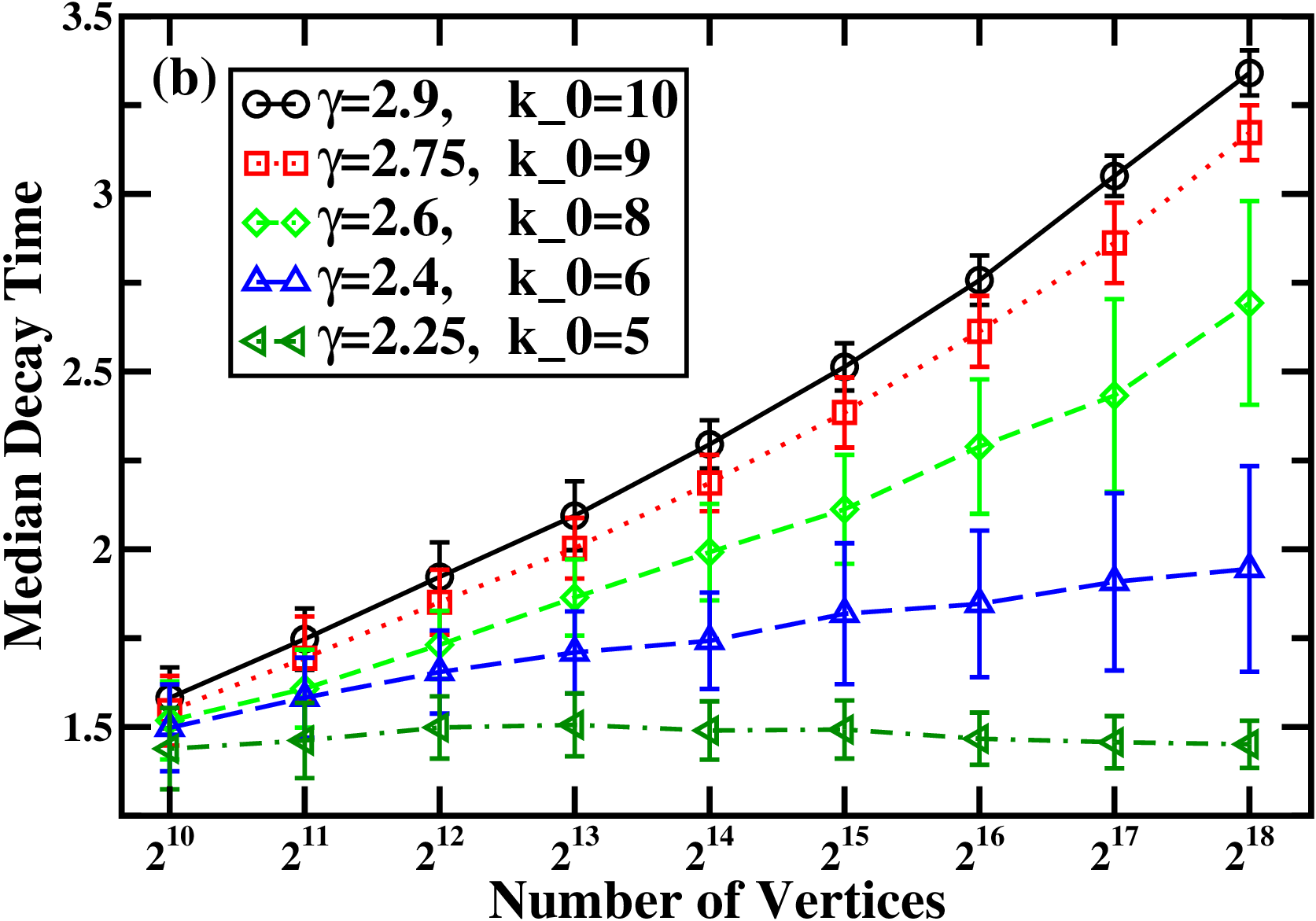}
  \caption{
    \label{fig:QA_SRN_s}
    Median decay time as a function of vertex size $N$ for random  simple-networks and
    random sequential or asychronous updating. Each data point is obtained by performing
    simulations for the same ensemble of
    $100$  simple-networks as in Fig.~\ref{fig:QA_SRN_p}. 
    (a) Median decay time needed for $|Q(t)-1/2|$ to exceed $1/4$ for the
    first time; and (b) Median decay time needed for $|\Lambda(t)|$ to exceed $1/2$
    for the first time. 
  }
\end{figure}

We use rule B as described in Sec.~\ref{sec:model_network} in order to 
generate an ensemble of scale-free random simple-networks. Each vertex of the network is occupied by a binary variable or spin which evolves again according to local majority rule dynamics. 
We use both parallel and random sequential updating starting from random initial
conditions as specified by Eq.~(\ref{eq:zero_AQ}).
The degree sequences of the generated networks are the same as those used in the preceding
section. Therefore, if the results of this subsection are different from those of 
Sec.~\ref{sec:simulation_MRN}, the differences must be related to  the correlations in the random
simple-networks. 
Our simulational results are summarized in
Fig.~\ref{fig:QA_SRN_p} and Fig.~\ref{fig:QA_SRN_s}  for parallel and 
random sequential updating, respectively.

In Fig.~\ref{fig:QA_SRN_p}(a) and Fig.~\ref{fig:QA_SRN_s}(a), we display 
 the median decay time $t_{d,m}$ needed for the network to reach an activity pattern
 that is characterized by a $nn$--spin up probability $Q(t)$ that satisfies
$|Q(t)-Q(0)| \geq 1/4$ when the initial pattern is characterized by the relation  (\ref{eq:zero_AQ}). 
Likewise, in  Fig.~\ref{fig:QA_SRN_p}(b)
and Fig.~\ref{fig:QA_SRN_s}(b), we  show the median decay time for the network to reach 
a pattern with average spin value
$|\Lambda(t)| \geq 1/2$ starting again from an initial pattern which satisfies
(\ref{eq:zero_AQ}).

For scaling exponent $\gamma > 5/2$,
the median decay time $t_{d,m}$ increases  with network size $N$ as  $t_{d,m} \sim (\ln N)^\eta$
with a growth exponent
$\eta$ that again appears to be slightly larger than the mean field value $\eta = 1$. The same behavior was observed in the preceding 
section for random multi-networks. For $\gamma = 2.25$, which is smaller than but not close to $\gamma = 5/2$, 
the median decay times are essentially independent of   the network size $N$. The latter behavior was also found for random sequential updating of multi-networks.  For $\gamma = 2.4$, i.e.,   smaller than but close to $\gamma = 5/2$, on the other hand, both
Fig.~\ref{fig:QA_SRN_p} and Fig.~\ref{fig:QA_SRN_s} demonstrate
that  the median decay time first increases slowly and then saturates for increasing network size $N$.  In other words, the function $t_{d,m} = t_{d,m}( \ln N)$ is now convex downwards. 
This situation is remarkably different from what was found  in  random
multi-networks, see Fig.~\ref{fig:QA_MRN_p} and Fig.~\ref{fig:QA_MRN_s}, for which the 
function $t_{d,m} = t_{d,m}( \ln N)$ is   convex upwards over the whole range of accessible
$N$--values. 
Therefore,  for random simple-networks, we conclude that 
the median decay time $t_{d,m}$ becomes independent of network size $N$ in the limit of 
large $N$ if the scaling  exponent $\gamma$ of the scale-free degree distribution satisfies $\gamma < 5/2$ as predicted by mean field theory.

\section{Effective scaling exponent for multi-networks}
\label{sec:effective_scaling}

\begin{figure}[h]
  \includegraphics[width=0.6\linewidth]{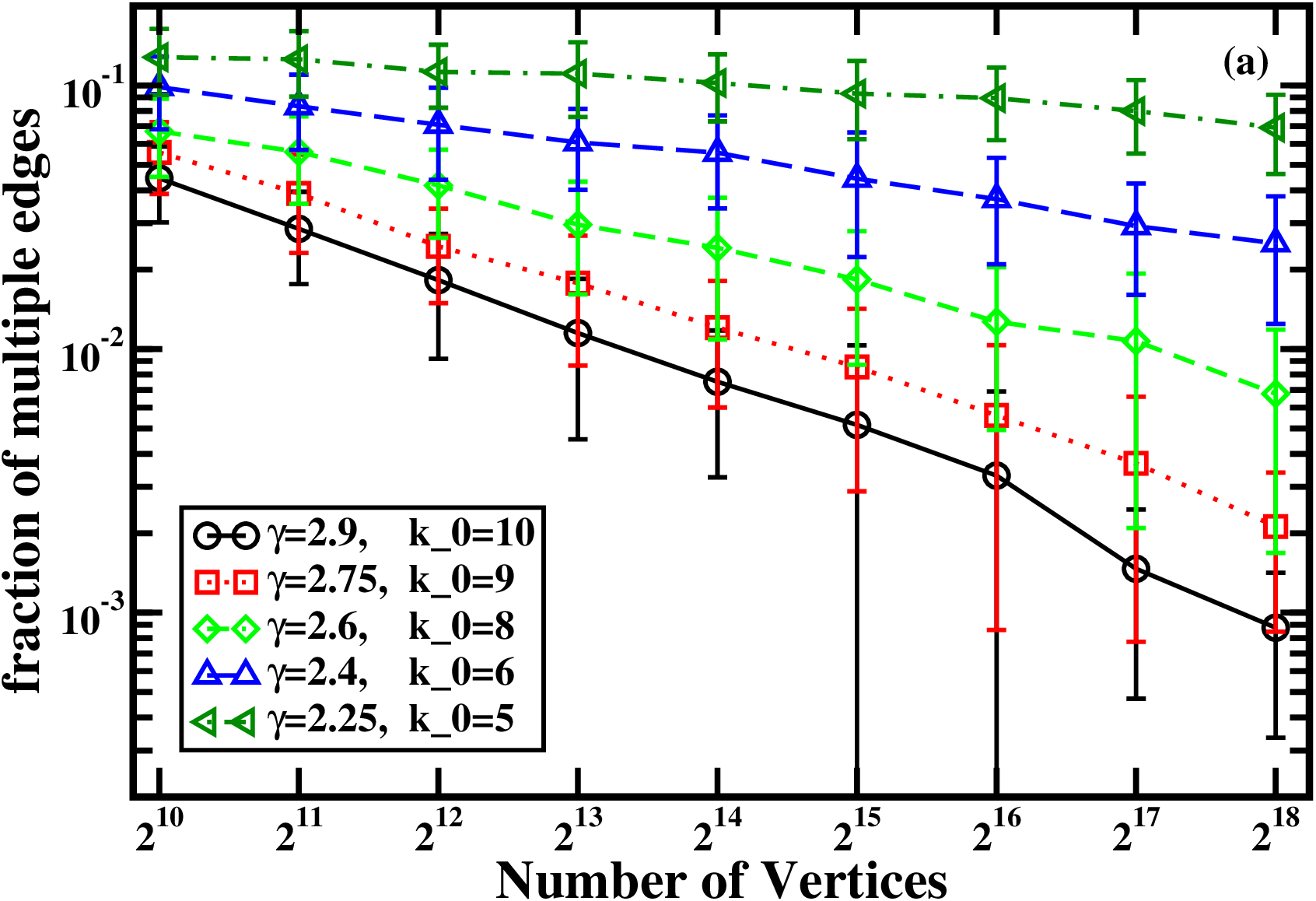}

  \vskip 0.5cm
  \includegraphics[width=0.6\linewidth]{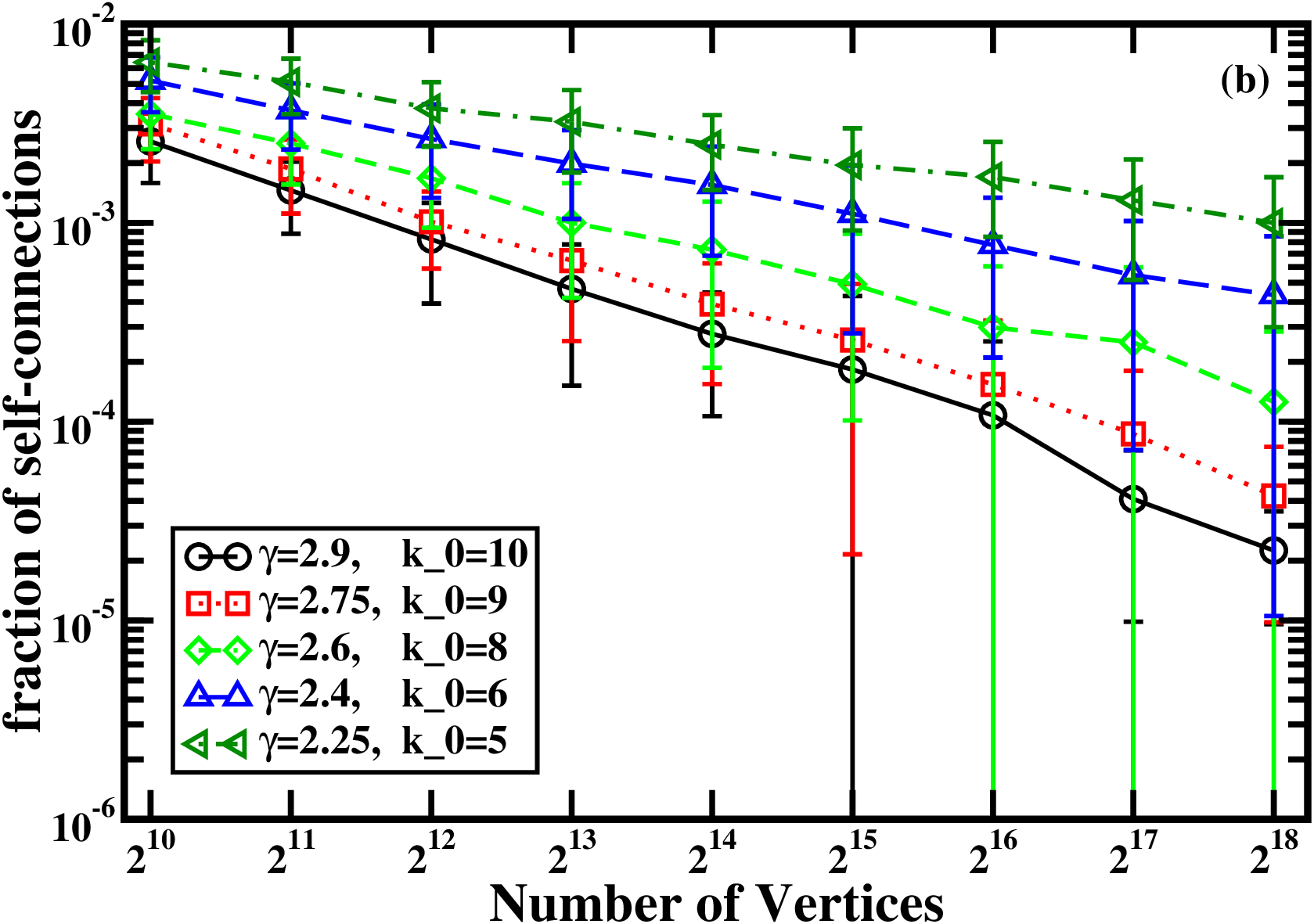}
  \caption{\label{fig:multi-self} 
    Fraction of (a) multiple edges   and (b)
    self-connections    in 
    a random scale-free multi-network. Each data point is an average over
    the same ensemble of $100$ random networks as used in the simulations of the preceding
    sections.
  }
\end{figure}

Multi-network  typically contain self-connections and multiple edges. As shown in 
Fig.~\ref{fig:multi-self}(a), the fraction of  edges that represent multiple edges is found 
to vary between about $10^{-1}$ and $10^{-3}$ depending on the number $N$ of vertices and 
the scaling exponent $\gamma$ of the scale-free degree distribution. Furthermore, the 
fraction of edges that represent self-connections varies between $10^{-2}$ and $10^{-5}$, 
see Fig.~\ref{fig:multi-self}(b). Both fractions decrease with increasing network size $N$ 
and increase with decreasing scaling exponent $\gamma$. This behavior can be understood 
from the following considerations. 

In a random multi-network, the probability that an edge is a self-connection of vertex $i$ is
equal to $\bigl( k_i / \sum_j k_j \bigr)^2 = k_i^2 / (4 M^2)$, where $k_i$ is the
degree of vertex $i$ and $M$ is the total number of edges as before. 
The average number of self-connections of  a certain vertex $i$ with degree $k_i$ is then
given by $ k_i^2/ 4M$. 
Therefore, the total number of self-connections is estimated to be
\begin{eqnarray}
  M_{\rm self} &= & \sum\limits_{i=1}^{N} { k_i^2 \over 4 M} 
  = \sum\limits_{k} N P(k) { k^2 \over 4 M}
  \label{eq:self-edges1} \\
  &\approx&{ k_0^2 (\gamma-1) \over 2 \langle k \rangle (3 - \gamma)} 
  { N^{(3-\gamma)/(\gamma-1)}-1 \over 1 - 1/N} 
  \label{eq:self-edges2}
\end{eqnarray}
for a scale-free network with scaling exponent $\gamma$. This relation  
shows that, for $\gamma \geq 3$, the total number of self-connetions
is of order unity. In contrast, for scale-free networks with $2 < \gamma <3$, the total 
number $M_{\rm self} $ of self-connections scales with
network size $N$ as $M_{\rm self} \sim N^{(3-\gamma)/(\gamma-1)}$ for large $N$. Therefore, 
the fraction of self-connections  decays as $M_{\rm self}/M \sim N^{(4-2 \gamma)/(\gamma-1)}$
with increasing network size $N$. 

Likewise, in a random multi-network, 
the  probability that an edge is formed between a certain  vertex $i$ with degree $k_i$ and another
vertex $j$ with degree
$k_j$  is given by 
\be
2 { k_i \over \sum_l k_l} { k_j \over \sum_l k_l} = { k_i k_j \over 2 M^2} \ .
\ee
Using this probabilility, one may derive an approximate expression for the total number $M_{mult}$ 
of multiple edges. 
The formula is a little bit more complex than Eq.~(\ref{eq:self-edges2}) so we
do not write down the explicit formula here. 
The average number of multiple edges 
is also found to be of order unity for  $\gamma \geq 3$ and to increase 
with network
size $N$ as a power-law for $2 < \gamma < 3$. This is consistent with the 
data shown in Fig.~\ref{fig:multi-self}(a).

\begin{figure}[h]
  \includegraphics[width=0.6\linewidth]{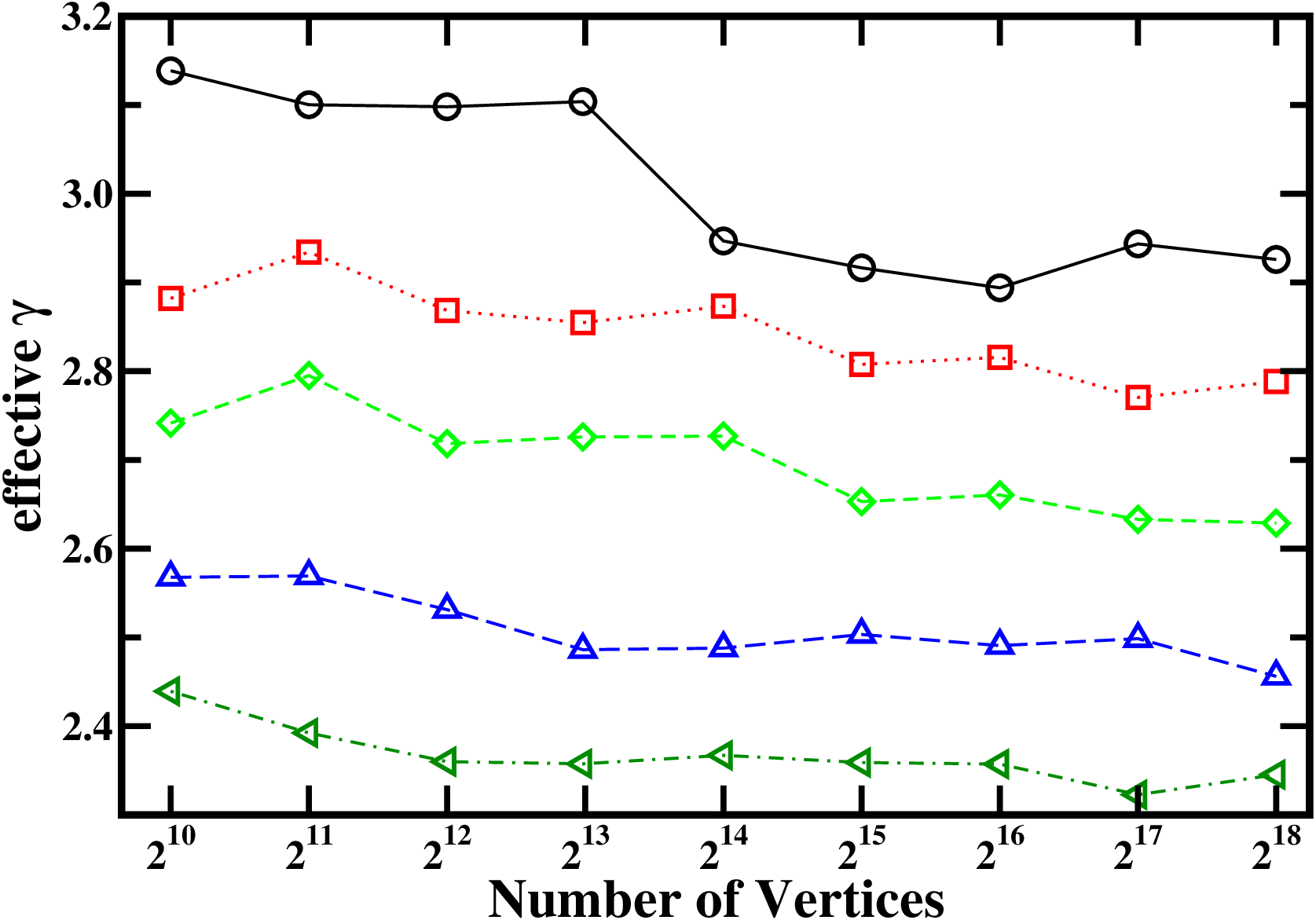}
  \caption{\label{fig:effectivegamma} 
    Effective   scaling exponent $\gamma_{\rm eff}$ for random simple-networks
    that have been obtained  by removal of 
   all multiple edges and
    self-connections from an ensemble of random multi-networks. Each data point is obtained by 
    merging all the $100$ vertex degree sequences corresponding to the
    $100$ multi-networks used in the simulation. The   degree distributions 
    of these multi-networks are characterized by the parameter values
    $\gamma=2.25$, $k_0=5$ (triangles left), $\gamma=2.4$, $k_0=6$
    (triangles up), $\gamma=2.6$, $k_0=8$ (diamonds), 
    $\gamma=2.75$, $k_0=9$ (squares), and $\gamma=2.9$, $k_0=10$
    (circles).
  }
\end{figure}

Furthermore, we find that the removal of all self-connections and multiple edges from 
a scale-free multi-network leads to  a simple-network that is again scale free but with 
the modified vertex degree distribution 
\be 
  \label{eq:effective}
  P_{\rm eff} (k) \sim k^{-\gamma_{\rm eff}}  
  \quad {\rm with} \quad
\gamma_{\rm eff} = \gamma_{\rm eff} (\gamma, N)    \, .
\ee
In Eq.~(\ref{eq:effective}), $k$ is the effective degree of a vertex, i.e., the
number of connections of the vertex after all its self-connections are removed and all its
multiple edges are reduced to a single edge. The scaling exponent $\gamma_{\rm eff}$ can be obtained 
by calculating the cumulative distribution function of the effective degree $k$ 
\cite{Newman-2005}.
The functional dependence of  the effective  
scaling exponent $\gamma_{\rm eff}$ on the scaling exponent $\gamma$ of the 
original multi-network and on the network size
$N$ is displayed in Fig.~\ref{fig:effectivegamma}. 
Inspection of this figure shows  that the effective scaling exponent satisfies  $\gamma_{\rm eff} \geq \gamma$ and  
decreases towards $\gamma$
with increasing network size $N$. For $\gamma = 2.9, 2.75$,  and 2.6, the data strongly 
indicate that the effective exponent $\gamma_{\rm eff}$ becomes asymptotically equal to 
$\gamma$ for large $N$. We expect that the same behavior applies to the multi-networks with 
$\gamma < 5/2$ but this remains to be shown. 

Now, let us focus on the case $\gamma = 2.4$ for which we found some qualitative differences 
between    multi-  and   simple-networks. As shown in Fig.~\ref{fig:effectivegamma}, 
the effective scaling exponent is found to satisfy  $\gamma_{\rm eff} \simg 2.5$ up to $N = 2^{17}$
for   multi-networks with $\gamma = 2.4$.  Thus, these multi-networks lead  to 
simple-networks with $\gamma_{\rm eff} \geq 5/2$ for almost all values of $N$ that are   accessible 
to the simulations. In this way, we obtain 
a rather intuitive explanation for our difficulty
to confirm the mean field predictions for multi-networks with $\gamma = 2.4$.

As far as the  network's topology is concerned, both self-connections and multiple edges can be
regarded as redundant. However,     these self-connections and multiple edges
have very significant effects for the dynamics of activity patterns of these networks. 
Let us consider the following simple example. Suppose
there are two high-degree vertices $i$ and $j$ with initial spin state $\sigma_i=1$ and $\sigma_j=-1$.
Each of these two vertices is taken to have many self-connections, and there are many edges
between them. The self-connections tend to stabilize the current spin state of each vertex,
while the edges between $i$ and $j$ tend to align the two spins. The frustration
caused by the competition of these two effects can lead to an increase in the relaxation
time. More generally, in a scale-free multiple network with $\gamma < 3$, there exists a 
densly connected core of
high-degree vertices. Each vertex of this core has many self-connections (which tend to
stabilize its current spin state) and is connected by multiple edges to many vertices in this core
(which tend to align the spins on  neighboring vertices).
Starting from a random initial spin configuration, spin state frustrations may easily build up 
within such a core. Such frustrations can be `annealed' by  external influences as provided, e.g., 
by the interactions with vertices outside the core, but for scale-free multi-networks with $\gamma \sim 5/2$, this annealing process
may take a relatively long time, since the fraction of  `redundant' edges is high and the
effective scaling exponent is larger than $5/2$. 

In our simulation, we have studied only the dynamic properties of random scale-free networks. 
In general, the vertex degrees of neighboring vertices may be correlated.
In the case of simple networks, these correlations
are caused by prohibition of self-connections and multiple edges
\cite{Maslov-Sneppen-2002,Maslov-etal-2002,Boguna-etal-2004,Catanzaro-etal-2005}; 
in the case of multiple networks,
these correlations exist between the effective vertex degrees of neighboring vertices. 
Such correlations of vertex degrees may also have some influence on the majority-rule dynamics. 
In order to study  the influence of vertex degree correlations on the dynamic network
processes in a quantitative manner,  one may generate ensembles of random networks with specified
vertex degree and specified vertex correlation patterns (see, e.g., \cite{Maslov-etal-2002}). 
In fact, the co-evolution of   correlations in the network structure 
and network dynamics represents a rather interesting topic  for future studies.

\section{Conclusion and discussion}
\label{sec:conclusion}

We have studied the time evolution of activity   patterns on scale-free random networks
which are characterized by the degree distribution $P(k) \sim 1/k^\gamma$ for $k_0 < k < 
k_{max}$. We constructed 
two different  ensembles of such networks as explained in Section \ref{sec:model_network}: 
(A) multi-networks  with  self-connections and multiple edges as generated by the configuration 
method  and (B) simple-networks without any self-connections and multiple edges as 
generated  by a combination of the configuration method and  an edge switching or   reshuffling algorithm.  We focussed on networks with minimal vertex degree $k_0 \geq 5$ for which 
all network graphs were found to consist of only a single component.

On each vertex of these networks, we   place a  binary variable or Ising-like spin 
$\sigma_i = \pm 1$.  For a given network with $N$ vertices, the activity pattern at time $t$ is  
described by the  spin configuration  $\{ \sigma (t) \} \equiv \{ \sigma_1(t), \sigma_2(t), \dots, \sigma_N(t) \}$.  The time evolution of this  pattern is govened by   local majority dynamics, 
which represents a Markov process in pattern space and is equivalent  to Glauber dynamics 
at zero temperature. We used two different updating schemes as explained
in Section \ref{sec:model_majority_rule}: (i) parallel or synchronous updating as defined by (\ref{MajorityRuleParallel}) and (ii) random sequential or 
asynchronous updating as in (\ref{MajorityRuleSequential}).

We focussed on the relaxation or decay of initial activity  patterns that are strongly disordered  
   both with respect to  the average spin value $\Lambda = \langle \sigma \rangle$ as given by 
 (\ref{eq:Lambda}) and with 
respect to the $nn$--spin up probability $Q$ as defined in (\ref{Qkt})  (where $nn$ stands for `nearest neighbor').  Indeed, these initial patterns are chosen to satisfy $\Lambda = 0$ and $Q = 1/2$ as in
(\ref{StronglyDisorderedPattern}). The relaxation process is measured by the decay time $t_d$  it takes to  reach a pattern with  $|\Lambda| \geq 1/2$ and/or $|Q - 1/2| \geq 1/4$. This decay time is governed by 
a single peak distribution from which we determined the median decay time $t_{d,m}$.

The mean field theory described in Section \ref{sec:analytic} predicts that the  $N$--dependence of the decay time  is  different for 
scaling exponents $\gamma < 5/2$ and $\gamma >  5/2$. For $\gamma < 5/2$, 
the typical decay time remains  finite even in the limit of large 
network size $N$. 
In contrast, this time scale increases as $\ln N$ with increasing $N$ for $\gamma >  5/2$.
In order to check these mean field predictions, we have
performed four types of  numerical simulations corresponding to parallel and random 
sequential updating of multi- and simple-networks. 

The simulation  data for  parallel updating of multi-networks are shown in 
Fig.~\ref{fig:QA_MRN_p}. In this case,  the  median decay time is found to  increase  with 
network size $N$
up to   $N =  2^{18}$  for all values of $\gamma$. Random sequential updating of the 
same ensemble of multi-networks leads to the data shown in  Fig.~\ref{fig:QA_MRN_s}. 
Inspection of this latter figure shows that 
the median decay time is now essentially independent of $N$ for $\gamma = 2.25$ 
as predicted by mean field theory. However, for $\gamma = 2.4$, this time scale still 
seems to increase with $N$ in contrast to the mean field predictions. 
This apparent increase can be understood if one defines an effective scaling exponent 
$\gamma_{\rm eff}$ for multi-networks as discussed in Section \ref{sec:effective_scaling}. 
As shown in Fig.~\ref{fig:effectivegamma}, this effective scaling exponent  decreases monotonically with increasing network  size $N$ and satisfies  $\gamma_{\rm eff} \geq \gamma$.  In particular, for 
multi-networks with  $\gamma = 2.4$, we find  $\gamma_{\rm eff} \simg 5/2$
up to $N = 2^{17}$. 

The simulation data for parallel and random sequential updating of simple-networks are
shown in Fig.~\ref{fig:QA_SRN_p} and Fig.~\ref{fig:QA_SRN_s}, respectively. In this case, 
both updating schemes are in complete agreement with the mean field predictions. 
Thus, we conclude that mean field theory is valid for the ensemble of random simple-networks
but is more difficult to confirm for the ensemble of random multi-networks because of 
strong finite size effects as described by the $N$--dependent value of the effective scaling
exponent $\gamma_{\rm eff}$. 

In the present article, we have analyzed the time evolution of 
strongly disordered    patterns at $t = 0$ into  partially  ordered patterns characterized by  
$|\Lambda| > 1/2$ and/or $|Q-1/2| > 1/4$ at $t = t_d$. These partially ordered patterns further evolve
towards completely ordered patterns. If the network has $\kappa$ components, one has 
$2^\kappa$ such ordered patterns. However,  there
is always a small but non-zero probability that the local majority rule dynamics does not
achieve complete order \cite{Castellano-PastorSatorras-2006}. 
One major reason for this failure is the existence of small network
components with certain `balanced' shapes that lead to domain boundaries and 
blinkers  \cite{Svenson-2002,Haggstrom-2002}.
The  probability to find such components is  expected to decrease with increasing
mean vertex degree $\langle k \rangle$ and increasing  minimal vertex degree $k_0$. 

Both for  $d$-dimensional
regular lattices \cite{Spirin-etal-2001} and for tree-like networks \cite{Melin-etal-1996}, 
the time evolution of activity patterns typically leads to  metastable states with many
domain boundaries and blinkers. In contrast, 
we did not find such a behavior for   random scale-free networks 
with $k_0 \geq 2$ as studied here. Indeed, 
our simulations show that  most  of the  spins are already  aligned 
after about 30 time steps, see Fig.~\ref{fig:QA_change}.

Our mean field analysis can also be applied to  more complicated dynamical models
as we have previously discussed in  \cite{Zhou-Lipowsky-2005}. 
One such extension is provided by   finite-connectivity Hopfield models \cite{Hopfield-1982}
 on  scale-free random networks. In this case, our 
mean field theory predicts  the storage capacity to grow as $N^\alpha$  with network size $N$ 
where the growth exponent $\alpha$ satisfies $1 >  \alpha > 0 $ provided  $2< \gamma < 5/2$.  
According
to   recent experimental studies \cite{Chialvo-2004,Eguiluz-etal-2005}, the functional networks
of the human brain are scale-free with a scaling exponent $\gamma \simeq 2.1$. In the latter case, 
our mean field theory leads to a storage capacity that grows as $N^\alpha$ with $\alpha \simeq 0.73$, 
i.e., almost as fast as for the original Hopfield models on complete graphs. 

Likewise, we have extended our mean field theory to  scale-free networks for which each vertex $i$
contains a  Potts-like variable  
$\sigma_i$  that can attain $q \geq 3$ values.  These dynamic systems exhibit $q$
ordered patterns (in each network component) and may evolve towards any of those patterns when they start from
a strongly disordered one. For these Potts-like systems, we find the same
distinction between the relaxation behavior for scale-free networks with $\gamma < 5/2$ and $\gamma > 5/2$ as in the case of the Ising-like systems studied here. We   also considered  some hybrid dynamics 
constructed from 
local majority dynamics and random Boolean dynamics: each spin variable $\sigma_i$  is updated 
with probability ${\cal P}$ according to  local majority rule dynamics  and with 
probability $1 - {\cal P}$ according to random Boolean dynamics as in the Kauffman models \cite{Kauffman-1969,Derrida-Pomeau-1986}.  
As long as ${\cal P} >0$, the relaxation behavior of this
more general class of models shows the same qualitative change at $\gamma=5/2$.
Finally, we have extended our mean field theory to binary or Ising-like variables on {\em directed}   
networks, see \cite{Zhou-Lipowsky-2005}. The elucidation  of these mean field results by 
appropriate simulation studies is  highly desirable and  remains to be done. 

\section{Appendix: Upper cut-off for vertex degree}

In this appendix, we explain the dependence of the maximal degree $k_{\rm max}^{(2)}$ on $N$ in more detail.
In Ref.~\cite{Cohen-etal-2000}, the scaling $k_{\rm max}^{(2)} \sim N^{{1\over \gamma-1}}$ was obtained from the
requirement that the average number of vertices with degree $k \geq k_{\max}^{(2)}$ should be of order
one, that is
\begin{equation}
  \label{eq:Cohen}
  N { \sum\limits_{k=k_{\rm max}^{(2)}}^\infty k^{-\gamma} \over \sum\limits_{k=k_0}^\infty k^{-\gamma}} \simeq 1 \ .
\end{equation}
It is instructive to rederive the same scaling relation for $k_{\rm max}^{(2)}$ through an alternative way. 
To do this, we  start from the scale-free degree distribution $P_{\infty}(k)$
which has the same form as that of $P(k)$ in Eq.~(\ref{eq:sf_degree_profile_2})
but with $k_{\max}=\infty$. The distribution $P_\infty(k)$  is used to generate $N$ random numbers 
$x_i$ with $i = 1, 2, \ldots, N$ and $x_i \geq  k_0$ which correspond to the vertex degrees 
of the $N$ vertices. Since the normalization factor ${\cal A}$ as given by Eq.~(\ref{eq:norm_A})
behaves as ${\cal A} \approx  k_0^{1-\gamma} /(\gamma-1)$ for large $k_{\rm max}$, the random variables 
$x_i$ are generated according to the probability density
\begin{equation}
  \label{eq:P_infty}
  P_\infty(x)  = (\gamma-1) k_0^{\gamma-1} x^{-\gamma} \ .
\end{equation}
The maximal value $x_{\rm max}$ of the $N$ random numbers $x_i$ is then governed by
the probability density
\begin{equation}
  \label{eq:rho_max}
  \rho( x_{\rm max} ) = N (\gamma - 1) k_0^{\gamma-1} x_{\rm max}^{-\gamma} 
  \Bigl( 1- \bigl( {k_0 \over x_{\rm max}} \bigr)^{\gamma-1} \Bigr)^{N-1} \ .
\end{equation}
It follows from this latter probability density, which is normalized as well,
that $x_{\rm max}$ has the average value
\begin{equation}
  \label{eq:x_max}
  \langle x_{\rm max} \rangle  = N k_0 B\bigl( {\gamma-2 \over \gamma-1},  N \bigr) \ ,
\end{equation}
where $B(z,N)$ is the standard beta function \cite{Abramowitz-Stegun-1972}. 
In the limit of large network
size $N$, one has $B(z,N) \approx \Gamma(z) \exp(z) N^{-z}$, which implies
\begin{equation}
  \label{eq:x_max_app}
  \langle x_{\rm max}\rangle  \approx  k_0 N^{1/(\gamma-1)} \ , 
\end{equation}
i.e., the same $k_0$ and $N$ dependence as for $k_{\rm max}^{(2)}$. The same
dependencies are also obtained for the most probable value of $x_{\rm max}$ which
corresponds to the maximum of the distribution $\rho(x_{\rm max})$ and is given by
\begin{equation}
  \label{x_max_mp}
  x_{\rm max}^{(mp)} = k_0 \bigl( {1 \over \gamma} + { \gamma-1\over \gamma} N \bigr)^{{1 \over \gamma-1}} \ .
\end{equation}

Using the probability density $\rho( x_{\rm max} )$ as given by (58), we obtain for 
the second moment of $x_{\rm max}$ the expression
\begin{equation}
  \label{eq:xmax2}
  \langle x_{\rm max}^2 \rangle = N k_0^2 B({\gamma-3\over \gamma-1}, N)
\end{equation}
For $\gamma < 3$, 
$\langle x_{\rm max}^2\rangle$ is infinite since one of the arguments of the beta function
in Eq.~(\ref{eq:xmax2}) is negative. 
Thus,  the fluctuations in $x_{\rm max}$ are unbounded
unless we use the natural upper cut-off  $k_{\rm max}^{(1)}=N-1$.

\section*{Acknowledgement}

We thank Kang Li, J\"org Menche, and Zhen Shao for helpful discussions.
The numerical simulations  were performed on the PC clusters of the  State     
Key Laboratory of Scientific and Engineering Computing, the Chinese Academy 
of Sciences.


\begin{thebibliography}{35}
\expandafter\ifx\csname natexlab\endcsname\relax\def\natexlab#1{#1}\fi
\expandafter\ifx\csname bibnamefont\endcsname\relax
  \def\bibnamefont#1{#1}\fi
\expandafter\ifx\csname bibfnamefont\endcsname\relax
  \def\bibfnamefont#1{#1}\fi
\expandafter\ifx\csname citenamefont\endcsname\relax
  \def\citenamefont#1{#1}\fi
\expandafter\ifx\csname url\endcsname\relax
  \def\url#1{\texttt{#1}}\fi
\expandafter\ifx\csname urlprefix\endcsname\relax\def\urlprefix{URL }\fi
\providecommand{\bibinfo}[2]{#2}
\providecommand{\eprint}[2][]{\url{#2}}

\bibitem[{\citenamefont{Albert and
  Barab{\'{a}}si}(2002)}]{Albert-Barabasi-2002}
\bibinfo{author}{\bibfnamefont{R.}~\bibnamefont{Albert}} \bibnamefont{and}
  \bibinfo{author}{\bibfnamefont{A.-L.} \bibnamefont{Barab{\'{a}}si}},
  \bibinfo{journal}{Rev. Mod. Phys.} \textbf{\bibinfo{volume}{74}},
  \bibinfo{pages}{47} (\bibinfo{year}{2002}).

\bibitem[{\citenamefont{Dorogovtsev and
  Mendes}(2002)}]{Dorogovtsev-Mendes-2002}
\bibinfo{author}{\bibfnamefont{S.~N.} \bibnamefont{Dorogovtsev}}
  \bibnamefont{and} \bibinfo{author}{\bibfnamefont{J.~F.~F.}
  \bibnamefont{Mendes}}, \bibinfo{journal}{Adv. Phys.}
  \textbf{\bibinfo{volume}{51}}, \bibinfo{pages}{1079} (\bibinfo{year}{2002}).

\bibitem[{\citenamefont{Newman}(2003)}]{Newman-2003}
\bibinfo{author}{\bibfnamefont{M.~E.~J.} \bibnamefont{Newman}},
  \bibinfo{journal}{SIAM Rev.} \textbf{\bibinfo{volume}{45}},
  \bibinfo{pages}{167} (\bibinfo{year}{2003}).

\bibitem[{\citenamefont{Barab{\'{a}}si and
  Albert}(1999)}]{Barabasi-Albert-1999}
\bibinfo{author}{\bibfnamefont{A.-L.} \bibnamefont{Barab{\'{a}}si}}
  \bibnamefont{and} \bibinfo{author}{\bibfnamefont{R.}~\bibnamefont{Albert}},
  \bibinfo{journal}{Science} \textbf{\bibinfo{volume}{286}},
  \bibinfo{pages}{509} (\bibinfo{year}{1999}).

\bibitem[{\citenamefont{Valverde et~al.}(2002)\citenamefont{Valverde, {Ferrer
  Cancho}, and Sol{\'{e}}}}]{Valverde-etal-2002}
\bibinfo{author}{\bibfnamefont{S.}~\bibnamefont{Valverde}},
  \bibinfo{author}{\bibfnamefont{R.}~\bibnamefont{{Ferrer Cancho}}},
  \bibnamefont{and} \bibinfo{author}{\bibfnamefont{R.~V.}
  \bibnamefont{Sol{\'{e}}}}, \bibinfo{journal}{Europhys. Lett.}
  \textbf{\bibinfo{volume}{60}}, \bibinfo{pages}{512} (\bibinfo{year}{2002}).

\bibitem[{\citenamefont{{Ferrer i Cancho} and
  Sol{\'{e}}}(2003)}]{FerrerICancho-Sole-2003}
\bibinfo{author}{\bibfnamefont{R.}~\bibnamefont{{Ferrer i Cancho}}}
  \bibnamefont{and} \bibinfo{author}{\bibfnamefont{R.~V.}
  \bibnamefont{Sol{\'{e}}}}, in \emph{\bibinfo{booktitle}{Lecture Notes in
  Physics 625: Statistical Mechanics of Complex Networks}}, edited by
  \bibinfo{editor}{\bibfnamefont{R.}~\bibnamefont{Pastor-Satorras}},
  \bibinfo{editor}{\bibfnamefont{M.}~\bibnamefont{Rubi}}, \bibnamefont{and}
  \bibinfo{editor}{\bibfnamefont{A.}~\bibnamefont{{Diaz-Guilera}}}
  (\bibinfo{publisher}{Springer}, \bibinfo{address}{Berlin},
  \bibinfo{year}{2003}), chap.~\bibinfo{chapter}{6}, pp.
  \bibinfo{pages}{114--126}.

\bibitem[{\citenamefont{Variano et~al.}(2004)\citenamefont{Variano, McCoy, and
  Lipson}}]{Variano-etal-2004}
\bibinfo{author}{\bibfnamefont{E.~A.} \bibnamefont{Variano}},
  \bibinfo{author}{\bibfnamefont{J.~H.} \bibnamefont{McCoy}}, \bibnamefont{and}
  \bibinfo{author}{\bibfnamefont{H.}~\bibnamefont{Lipson}},
  \bibinfo{journal}{Phys. Rev. Lett.} \textbf{\bibinfo{volume}{92}},
  \bibinfo{pages}{188701} (\bibinfo{year}{2004}).

\bibitem[{\citenamefont{Wang et~al.}(2005)\citenamefont{Wang, Wang, Hu, Yan,
  and Ou}}]{Wang-etal-2005}
\bibinfo{author}{\bibfnamefont{W.-X.} \bibnamefont{Wang}},
  \bibinfo{author}{\bibfnamefont{B.-H.} \bibnamefont{Wang}},
  \bibinfo{author}{\bibfnamefont{B.}~\bibnamefont{Hu}},
  \bibinfo{author}{\bibfnamefont{G.}~\bibnamefont{Yan}}, \bibnamefont{and}
  \bibinfo{author}{\bibfnamefont{Q.}~\bibnamefont{Ou}}, \bibinfo{journal}{Phys.
  Rev. Lett.} \textbf{\bibinfo{volume}{94}}, \bibinfo{pages}{188702}
  (\bibinfo{year}{2005}).

\bibitem[{\citenamefont{Zhou and Lipowsky}(2005)}]{Zhou-Lipowsky-2005}
\bibinfo{author}{\bibfnamefont{H.}~\bibnamefont{Zhou}} \bibnamefont{and}
  \bibinfo{author}{\bibfnamefont{R.}~\bibnamefont{Lipowsky}},
  \bibinfo{journal}{Proc. Natl. Acad. Sci. USA} \textbf{\bibinfo{volume}{102}},
  \bibinfo{pages}{10052} (\bibinfo{year}{2005}).

\bibitem[{\citenamefont{Aldana and Cluzel}(2003)}]{Aldana-Cluzel-2003}
\bibinfo{author}{\bibfnamefont{M.}~\bibnamefont{Aldana}} \bibnamefont{and}
  \bibinfo{author}{\bibfnamefont{P.}~\bibnamefont{Cluzel}},
  \bibinfo{journal}{Proc. Natl. Acad. Sci. USA} \textbf{\bibinfo{volume}{100}},
  \bibinfo{pages}{8710} (\bibinfo{year}{2003}).

\bibitem[{\citenamefont{Catanzaro et~al.}(2005)\citenamefont{Catanzaro,
  Bogu{\~{n}}{\'{a}}, and {Pastor-Satorras}}}]{Catanzaro-etal-2005}
\bibinfo{author}{\bibfnamefont{M.}~\bibnamefont{Catanzaro}},
  \bibinfo{author}{\bibfnamefont{M.}~\bibnamefont{Bogu{\~{n}}{\'{a}}}},
  \bibnamefont{and}
  \bibinfo{author}{\bibfnamefont{R.}~\bibnamefont{{Pastor-Satorras}}},
  \bibinfo{journal}{Phys. Rev. E} \textbf{\bibinfo{volume}{71}},
  \bibinfo{pages}{027103} (\bibinfo{year}{2005}).

\bibitem[{\citenamefont{Castellano and
  {Pastor-Satorras}}(2006)}]{Castellano-PastorSatorras-2006}
\bibinfo{author}{\bibfnamefont{C.}~\bibnamefont{Castellano}} \bibnamefont{and}
  \bibinfo{author}{\bibfnamefont{R.}~\bibnamefont{{Pastor-Satorras}}},
  \bibinfo{journal}{J. Stat. Mech.}, \bibinfo{pages}{P05001} (\bibinfo{year}{2006}).

\bibitem[{\citenamefont{Cohen et~al.}(2000)\citenamefont{Cohen, Erez,
  {ben-Avraham}, and Havlin}}]{Cohen-etal-2000}
\bibinfo{author}{\bibfnamefont{R.}~\bibnamefont{Cohen}},
  \bibinfo{author}{\bibfnamefont{K.}~\bibnamefont{Erez}},
  \bibinfo{author}{\bibfnamefont{D.}~\bibnamefont{{ben-Avraham}}},
  \bibnamefont{and} \bibinfo{author}{\bibfnamefont{S.}~\bibnamefont{Havlin}},
  \bibinfo{journal}{Phys. Rev. Lett.} \textbf{\bibinfo{volume}{85}},
  \bibinfo{pages}{4626} (\bibinfo{year}{2000}).

\bibitem[{\citenamefont{Milo et~al.}(2003)\citenamefont{Milo, Kashtan,
  Itzkovitz, Newman, and Alon}}]{Milo-etal-2003}
\bibinfo{author}{\bibfnamefont{R.}~\bibnamefont{Milo}},
  \bibinfo{author}{\bibfnamefont{N.}~\bibnamefont{Kashtan}},
  \bibinfo{author}{\bibfnamefont{S.}~\bibnamefont{Itzkovitz}},
  \bibinfo{author}{\bibfnamefont{M.~E.~J.} \bibnamefont{Newman}},
  \bibnamefont{and} \bibinfo{author}{\bibfnamefont{U.}~\bibnamefont{Alon}},
  \emph{\bibinfo{title}{On the uniform generation of random graphs with
  prescribed degree sequences}}, \bibinfo{howpublished}{e-print:
  cond-mat/0312028} (\bibinfo{year}{2003}).

\bibitem[{\citenamefont{Zhou}(2002)}]{Zhou-2002}
\bibinfo{author}{\bibfnamefont{H.}~\bibnamefont{Zhou}}, \bibinfo{journal}{Phys.
  Rev. E} \textbf{\bibinfo{volume}{66}}, \bibinfo{pages}{016125}
  (\bibinfo{year}{2002}).

\bibitem[{\citenamefont{Glauber}(1963)}]{Glauber-1963}
\bibinfo{author}{\bibfnamefont{R.~J.} \bibnamefont{Glauber}},
  \bibinfo{journal}{J. Math. Phys.} \textbf{\bibinfo{volume}{4}},
  \bibinfo{pages}{294} (\bibinfo{year}{1963}).

\bibitem[{\citenamefont{{Bar-Yam} and Epstein}(2004)}]{BarYam-Epstein-2004}
\bibinfo{author}{\bibfnamefont{Y.}~\bibnamefont{{Bar-Yam}}} \bibnamefont{and}
  \bibinfo{author}{\bibfnamefont{I.~R.} \bibnamefont{Epstein}},
  \bibinfo{journal}{Proc. Natl. Acad. Sci. USA} \textbf{\bibinfo{volume}{101}},
  \bibinfo{pages}{4341} (\bibinfo{year}{2004}).

\bibitem[{\citenamefont{Maslov and Sneppen}(2002)}]{Maslov-Sneppen-2002}
\bibinfo{author}{\bibfnamefont{S.}~\bibnamefont{Maslov}} \bibnamefont{and}
  \bibinfo{author}{\bibfnamefont{K.}~\bibnamefont{Sneppen}},
  \bibinfo{journal}{Science} \textbf{\bibinfo{volume}{296}},
  \bibinfo{pages}{910} (\bibinfo{year}{2002}).

\bibitem[{\citenamefont{Maslov et~al.}(2002)\citenamefont{Maslov, Sneppen, and
  Zaliznyak}}]{Maslov-etal-2002}
\bibinfo{author}{\bibfnamefont{S.}~\bibnamefont{Maslov}},
  \bibinfo{author}{\bibfnamefont{K.}~\bibnamefont{Sneppen}}, \bibnamefont{and}
  \bibinfo{author}{\bibfnamefont{A.}~\bibnamefont{Zaliznyak}},
  \bibinfo{journal}{Physica A} \textbf{\bibinfo{volume}{333}},
  \bibinfo{pages}{529} (\bibinfo{year}{2002}).

\bibitem[{\citenamefont{Bogu{\~{n}}{\'{a}}
  et~al.}(2004)\citenamefont{Bogu{\~{n}}{\'{a}}, {Pastor-Satorras}, and
  Vespignani}}]{Boguna-etal-2004}
\bibinfo{author}{\bibfnamefont{M.}~\bibnamefont{Bogu{\~{n}}{\'{a}}}},
  \bibinfo{author}{\bibfnamefont{R.}~\bibnamefont{{Pastor-Satorras}}},
  \bibnamefont{and}
  \bibinfo{author}{\bibfnamefont{A.}~\bibnamefont{Vespignani}},
  \bibinfo{journal}{Eur. Phys. J. B} \textbf{\bibinfo{volume}{38}},
  \bibinfo{pages}{205} (\bibinfo{year}{2004}).

\bibitem[{\citenamefont{Krapivsky and Redner}(2003)}]{Krapivsky-Redner-2003}
\bibinfo{author}{\bibfnamefont{P.~L.} \bibnamefont{Krapivsky}}
  \bibnamefont{and} \bibinfo{author}{\bibfnamefont{S.}~\bibnamefont{Redner}},
  \bibinfo{journal}{Phys. Rev. Lett.} \textbf{\bibinfo{volume}{90}},
  \bibinfo{pages}{238701} (\bibinfo{year}{2003}).

\bibitem[{\citenamefont{Aldana and Larralde}(2004)}]{Aldana-Larralde-2004}
\bibinfo{author}{\bibfnamefont{M.}~\bibnamefont{Aldana}} \bibnamefont{and}
  \bibinfo{author}{\bibfnamefont{H.}~\bibnamefont{Larralde}},
  \bibinfo{journal}{Phys. Rev. E} \textbf{\bibinfo{volume}{70}},
  \bibinfo{pages}{066130} (\bibinfo{year}{2004}).

\bibitem[{\citenamefont{Sood and Redner}(2005)}]{Sood-Redner-2005}
\bibinfo{author}{\bibfnamefont{V.}~\bibnamefont{Sood}} \bibnamefont{and}
  \bibinfo{author}{\bibfnamefont{S.}~\bibnamefont{Redner}},
  \bibinfo{journal}{Phys. Rev. Lett.} \textbf{\bibinfo{volume}{94}},
  \bibinfo{pages}{178701} (\bibinfo{year}{2005}).

\bibitem[{\citenamefont{Zhu et~al.}(2004)\citenamefont{Zhu, Xiong, Tian, Li,
  and Jiang}}]{Zhu-etal-2004}
\bibinfo{author}{\bibfnamefont{C.-P.} \bibnamefont{Zhu}},
  \bibinfo{author}{\bibfnamefont{S.-J.} \bibnamefont{Xiong}},
  \bibinfo{author}{\bibfnamefont{Y.-J.} \bibnamefont{Tian}},
  \bibinfo{author}{\bibfnamefont{N.}~\bibnamefont{Li}}, \bibnamefont{and}
  \bibinfo{author}{\bibfnamefont{K.-S.} \bibnamefont{Jiang}},
  \bibinfo{journal}{Phys. Rev. Lett.} \textbf{\bibinfo{volume}{92}},
  \bibinfo{pages}{218702} (\bibinfo{year}{2004}).

\bibitem[{\citenamefont{Newman}(2005)}]{Newman-2005}
\bibinfo{author}{\bibfnamefont{M.~E.~J.} \bibnamefont{Newman}},
  \bibinfo{journal}{Contemporary Phys.} \textbf{\bibinfo{volume}{46}},
  \bibinfo{pages}{323} (\bibinfo{year}{2005}).

\bibitem[{\citenamefont{Svenson}(2002)}]{Svenson-2002}
\bibinfo{author}{\bibfnamefont{P.}~\bibnamefont{Svenson}},
  \bibinfo{journal}{Phys. Rev. E} \textbf{\bibinfo{volume}{64}},
  \bibinfo{pages}{036122} (\bibinfo{year}{2002}).

\bibitem[{\citenamefont{H{\"{a}}ggstr{\"{o}}m}(2002)}]{Haggstrom-2002}
\bibinfo{author}{\bibfnamefont{O.}~\bibnamefont{H{\"{a}}ggstr{\"{o}}m}},
  \bibinfo{journal}{Physica A} \textbf{\bibinfo{volume}{310}},
  \bibinfo{pages}{275} (\bibinfo{year}{2002}).

\bibitem[{\citenamefont{Spirin et~al.}(2001)\citenamefont{Spirin, Krapivsky,
  and Redner}}]{Spirin-etal-2001}
\bibinfo{author}{\bibfnamefont{V.}~\bibnamefont{Spirin}},
  \bibinfo{author}{\bibfnamefont{P.~L.} \bibnamefont{Krapivsky}},
  \bibnamefont{and} \bibinfo{author}{\bibfnamefont{S.}~\bibnamefont{Redner}},
  \bibinfo{journal}{Phys. Rev. E} \textbf{\bibinfo{volume}{65}},
  \bibinfo{pages}{016119} (\bibinfo{year}{2001}).

\bibitem[{\citenamefont{M{\'{e}}lin et~al.}(1996)\citenamefont{M{\'{e}}lin,
  {Angl{\`{e}}s {d'}Auriac}, Chandra, and Doucot}}]{Melin-etal-1996}
\bibinfo{author}{\bibfnamefont{R.}~\bibnamefont{M{\'{e}}lin}},
  \bibinfo{author}{\bibfnamefont{J.~C.} \bibnamefont{{Angl{\`{e}}s
  {d'}Auriac}}}, \bibinfo{author}{\bibfnamefont{P.}~\bibnamefont{Chandra}},
  \bibnamefont{and} \bibinfo{author}{\bibfnamefont{B.}~\bibnamefont{Doucot}},
  \bibinfo{journal}{J. Phys. A: Math. Gen.} \textbf{\bibinfo{volume}{29}},
  \bibinfo{pages}{5773} (\bibinfo{year}{1996}).

\bibitem[{\citenamefont{Hopfield}(1982)}]{Hopfield-1982}
\bibinfo{author}{\bibfnamefont{J.~J.} \bibnamefont{Hopfield}},
  \bibinfo{journal}{Proc. Natl. Acad. Sci. USA} \textbf{\bibinfo{volume}{79}},
  \bibinfo{pages}{2554} (\bibinfo{year}{1982}).

\bibitem[{\citenamefont{Chialvo}(2004)}]{Chialvo-2004}
\bibinfo{author}{\bibfnamefont{D.~R.} \bibnamefont{Chialvo}},
  \bibinfo{journal}{Physica A} \textbf{\bibinfo{volume}{340}},
  \bibinfo{pages}{756} (\bibinfo{year}{2004}).

\bibitem[{\citenamefont{Eguiluz et~al.}(2005)\citenamefont{Eguiluz, Chialvo,
  Cecchi, Baliki, and Apkarian}}]{Eguiluz-etal-2005}
\bibinfo{author}{\bibfnamefont{V.~M.} \bibnamefont{Eguiluz}},
  \bibinfo{author}{\bibfnamefont{D.~R.} \bibnamefont{Chialvo}},
  \bibinfo{author}{\bibfnamefont{G.~A.} \bibnamefont{Cecchi}},
  \bibinfo{author}{\bibfnamefont{M.}~\bibnamefont{Baliki}}, \bibnamefont{and}
  \bibinfo{author}{\bibfnamefont{A.~V.} \bibnamefont{Apkarian}},
  \bibinfo{journal}{Phys. Rev. Lett.} \textbf{\bibinfo{volume}{94}},
  \bibinfo{pages}{018102} (\bibinfo{year}{2005}).

\bibitem[{\citenamefont{Kauffman}(1969)}]{Kauffman-1969}
\bibinfo{author}{\bibfnamefont{S.~A.} \bibnamefont{Kauffman}},
  \bibinfo{journal}{J. Theo. Biol.} \textbf{\bibinfo{volume}{22}},
  \bibinfo{pages}{437} (\bibinfo{year}{1969}).

\bibitem[{\citenamefont{Derrida and Pomeau}(1986)}]{Derrida-Pomeau-1986}
\bibinfo{author}{\bibfnamefont{B.}~\bibnamefont{Derrida}} \bibnamefont{and}
  \bibinfo{author}{\bibfnamefont{Y.}~\bibnamefont{Pomeau}},
  \bibinfo{journal}{Europhys. Lett.} \textbf{\bibinfo{volume}{1}},
  \bibinfo{pages}{45} (\bibinfo{year}{1986}).

\bibitem[{\citenamefont{Abramowitz and Stegun}(1972)}]{Abramowitz-Stegun-1972}
\bibinfo{author}{\bibfnamefont{M.}~\bibnamefont{Abramowitz}} \bibnamefont{and}
  \bibinfo{author}{\bibfnamefont{I.~A.} \bibnamefont{Stegun}},
  \emph{\bibinfo{title}{Handbook of mathematical functions}}
  (\bibinfo{publisher}{Dover Publications}, \bibinfo{address}{New York},
  \bibinfo{year}{1972}).

\end{thebibliography}

\end{document}